\documentclass[preprint,showpacs,amsfonts,aps]{revtex4}
\draft 
\usepackage{graphicx}
\usepackage{bbold}
\begin{document} 


\title{The classical Schr\"{o}dinger equation\footnote{Published version to appear in Pioneer Journal of Mathematical Physics and its Applications. This version ammended for subsequent publication of the physical interpretation due to the author.}}
\author{K. R. W. Jones}    
\date{2 March, 2014} 
\address{Sensors Divsion,\\
Alphaxon Research LLC,\\
PO Box 376, Milsons Point,\\
NSW 1565, Australia.}
\email{kingsley.jones@alphaxon.com}
\begin{abstract}
Using a simple geometrical construction based upon the linear action of
the Heisenberg--Weyl group we deduce a new nonlinear Schr\"{o}dinger
equation that provides an exact dynamic and energetic model of any classical
system whatsoever, be it integrable, nonintegrable or chaotic. Within our
model classical phase space points are represented by equivalence classes of
wavefunctions that have identical position and momentum expectation values.
Transport of these equivalence classes without dispersion leads to a system of wavefunction dynamics such that
the expectation values track classical trajectories {\em precisely\/} for arbitrarily long times. 
Interestingly, the value of $\hbar$ proves immaterial
for the purpose of constructing this alternative representation of classical
point mechanics. The new feature which $\hbar$ does mediate concerns a 
simple embedding of the quantum geometric phase within classical mechanics. 
We discuss problems of physical interpretation and discover a simple route to
recover the ordinary linear Schr\"{o}dinger equation.
\end{abstract}
\pacs{0230, 0365, 0545}
\maketitle

\newpage
\section{Introduction}
This paper describes a simple result that provides a direct way to relate classical dynamics to quantum dynamics as a particular approximation. The method involves the use of group theory to construct a second representation of classical point mechanics.
The result is a novel nonlinear integrodifferential wave equation which is dynamically and energetically equivalent to Hamiltonian
classical mechanics, but which makes direct use of wavefunctions. As such, it requires a different physical interpretation than
the ordinary linear Schr\"{o}dinger equation. Nevertheless, we believe it should prove useful in refining our understanding of the connection between the classical and quantum theories. With this  goal in mind, we first outline the question of an alternate representation for classical point mechanics. From that point, our argument proceeds in a precise and logical fashion.

One enduring puzzle of non-relativistic quantum theory is the role of the superposition principle in the classical limit. It is widely
supposed that quantum theory is universal and that the superposition principle is an exact property of the dynamics. This is in
seeming conflict with the properties of classical dynamics, which may in general be nonlinear.  This observation has motivated recurrent  interest in  nonlinear generalizations of the Schr\"{o}dinger equation\cite{debrog,bial}. Perhaps the most common
rationale is the idea that nonlinearity might suppress wavepacket dispersion via soliton solutions having properties closer to 
the familiar concept of a classical particle\cite{bial}. A related rationale concerns the current irreconcilable duality of quantum dynamics regarding the two distinct processes of unitary evolution and wavepacket collapse\cite{qmt}. In this connection, 
some suggest that the stochastic process of collapse might be subsumed as a possible nonlinear evolutionary process that 
is somehow characteristic of quantum  measurements\cite{pearle,bell}.

Whatever the rationale, the most common approach to the introduction of
quantal nonlinearity has been to consider the addition of mild perturbative
terms to the standard Schr\"{o}dinger equation. Perhaps the best known paper
of this variety is that by Bi\`{a}lynicki-Birula and Mycielski\cite{bial}.
Their study presented a particular kind of logarithmic nonlinearity as being
appropriate to the physical constraint that separated systems should combine
in a manner that avoids their quantal entanglement in the absence of a mutual
interaction. That work led to some careful empirical tests via neutron
interferometry which placed quite stringent upper bounds upon the strength of
the logarithmic quantal nonlinearity\cite{shull,klein}.
  
Weinberg\cite{wein1,wein2} has revived the topic via the elucidation
of a far more general theory of nonlinear quantal evolution. This theory is
distinguished by its use of the projective character of the quantum space of
states to extend the proper treatment of separated systems beyond the class
of purely logarithmic nonlinearities. Although the stated motivation remains
that of providing theoretical guidance to the precision tests of quantal
linearity, where the implicit expectation is that of a null result, the
theory itself is very elegant. So much so that one might believe in it were
it not for the fact that we have yet to observe quantal nonlinearity of any
kind in any system.

Given that the implications of the discovery of quantal nonlinearity would
be so profound, and noting that Weinberg's recent contribution actually is a
theory of quite general character, it is our belief that the current approach
to the topic of experiment is perhaps too negative. What one may need, if
nonlinearity is true physics, is some guidance about a new place to look
where there has not previously been an incentive to look. In this respect, the
close formal similarity between Weinberg's theory and Hamiltonian classical
mechanics suggests that the discussion of quantal nonlinearity might now
fruitfully be widened to include the larger question of the connection
between classical and quantum mechanics, and the curious possibility that
nonlinearity may actually be connected with the classical limit. In this
respect, it is significant that two notable precursors\cite{strocch,heslot}
of Weinberg's formalism, were guided by aspects of the formal similarity
between classical and quantal dynamics rather than an investigation into
possible quantal nonlinearities.

An alternative motivation for studying nonlinear quantum mechanics concerns
the apparent absence\cite{gutz} of dynamical chaos within the ordinary linear
theory of quantum mechanics. In stating this position we are influenced
considerably by the recent paper of Ford et al\cite{ford}. Given that
dynamical chaos is present within the nonlinear regime of the special theory
of classical mechanics, it becomes plausible  to suppose, for the sake of
exploration, that the border regions of the classical limit may well be just
that special place where quantal nonlinearity could become a strong rather
than a weak effect. Such an idea is heretical, but cannot be immediately
discounted. The reason why lies in the fact that an experimental probing of
systems that are neither manifestly quantal nor manifestly classical presents
peculiar problems of both definition and design within the current
orthodoxy. One prefers to have a clean system of definite type. Even to
imagine otherwise begs of a theoretical guide we do not as yet have, namely a
prototype mesoquantal mechanics. 

Returning now to the presence or absence of quantal nonlinearity, it is a
fact that all null tests performed to date\cite{tests} have been carried out
upon systems which {\em were already known\/} to be well described by linear
quantum theory. There exists therefore the slim but enticing possibility that
modern physics may have let something slip through its net. In the
exploration of this question one therefore desires a guiding principle that
would help locate a candidate mesoquantal theory. The obvious necessity is
that a theory of this kind must subsume both ordinary linear quantum theory 
and ordinary classical mechanics. As such we should expect it to remain a
theory of wavefunctions. Note that linear quantum theory already achieves
this aim in large part by showing that classical mechanics is a good
approximation in the limit $\hbar\rightarrow 0$. We must therefore add to the
above a reason for believing that the nature of things might be more subtle.
In this respect the apparent reality of classical dynamical chaos and the
general absence of macroscopic superpositions provide some minimal guidance
that one might legitimately pursue a nonlinear theory.

These few tenuous clues, coupled with the formal similarity between Weinberg's
generalized nonlinear evolution theory and classical mechanics, motivated the
author to look specifically at the theory of  classical mechanics as
one possible manifestation of a nonlinear quantum theory. Our
approach to this exercise was to pose the following simple question: 
\begin{quote}   
{\em Can one find an evolution equation for wavefunctions in Hilbert space
such that the expectation values of quantal position and momentum operators
will precisely follow the trajectories of an arbitrarily chosen classical
system?\/}        
\end{quote}  
Having posed this question we were somewhat surprised to find an affirmative
answer. The route we shall take leads to a general and seemingly very natural
nonlinear quantum evolution equation. The motivation of that equation arises
from the desire to construct a quantal model of classical mechanics. However,
its possible physical role transcends that special application.

Although Weinberg's theory acted primarily as a spur to our interest, it is
considered remarkable that the dynamical structure to which we are led seems
to share a close affinity with his. To the extent that Weinberg's theory
resembles Heisenberg's mechanics, our contribution resembles its natural
Schr\"{o}dinger analogue. Indeed, at the end of the paper we shall demonstrate
how knowledge of our own result enabled the author to find a precisely
analogous classical correspondence result within Weinberg's theory. Elsewhere,
we reported that argument as stemming from a plausible ansatz\cite{jones};
here one may discern its deeper origin.

In summary, the goal of this paper shall be the deduction of a new nonlinear
wave equation which models exact classical mechanics as the non--dispersive
evolution of wavefunctions in a manner that proves to be independent of
$\hbar$. Subsequent papers shall present further elaborations of this result
in relation to quantization theory, the classical limit and the possibility
of mesoquantal physics.

\section{Outline}
At heart our work depends upon a simple group theoretic result. We shall
merely replace the ordinary kinematical group of motions upon the phase
space of classical mechanics, the Abelian group of additive translations, by
its projective cousin, the Heisenberg--Weyl group\cite{weyl}. Everything else
follows from this single step. However, in order to make the argument widely
accessible we have included a significant amount of introductory material. 

Starting in \S {\rm III}, we review properties of the Heisenberg--Weyl group,
its multiplication, nonintegrable phase factor, action upon quantum states
and utility as a device for constructing a quantal analogue of classical
phase space. In \S {\rm IV} there appears an intuitive overview of our main
result. This we develop in \S {\rm V} and \S {\rm VI}. The outcome is a
nonlinear quantization prescription involving a $\psi$--dependent Hamiltonian
operator which then determines the desired wave equation. 

The most unfamiliar aspect of this  work shall likely be the fact that we seek
a quantization process which returns the classical dynamics one started with.
The idea seems odd to begin with, but will appear natural in the course of
our development. We shall gain via this route a translation of classical
mechanics directly into the mathematical framework of quantum mechanics. 
After \S {\rm VI}, the paper develops properties of the resulting dynamical
system, its peculiarities and its surprising connection with the ordinary
theory of linear quantum mechanics.

\section{Some Preliminary Mathematical Observations}
Our result follows from purely geometrical considerations of an elementary
nature. In essence we need only the following three sets of observations to
obtain the quantization prescription. All of the quoted results are standard.
Useful source materials include\cite{klau1,perel,zhang} for study of Weyl
operators and their relationship to coherent states, and\cite{arnold,mackey}
for mathematical aspects of classical and quantal dynamics.

Note that it proves easiest to deduce our results without explicit reference
to any particular Hilbert space representation. According to the Stone--von
Neumann theorem\cite{stone,voneu} all irreducible Hilbert space
representations of the canonical commutation relations are unitarily
equivalent\cite{prug}. It then follows that any calculation which one might
choose to carry out in a particular Hilbert space representation, but which
happens to be predicated purely upon the canonical commutation relations,
shall have a precise analogue in any other, for each such representation is
merely a special concrete manifestation of a single abstract mathematical
object. 

\subsection{Observations regarding Weyl operators}
We shall focus throughout upon a ubiquitous mathematical object known as the
Heisenberg--Weyl group\cite{perel,zhang}. Here we shall deal with quantal
systems having only one continuous degree of freedom. The generalization to
$n$ continuous degrees of freedom is straightforward\cite{perel}.

Let $\hat{a}^{\dagger}$ and $\hat{a}$ denote the usual creation and
annihilation operators satisfying the commutation relation
$[\hat{a},\hat{a}^{\dagger}]=\hat{I}$. In terms of these the Weyl operator,
$D(\alpha)$, is defined to be $$D(\alpha) \equiv\exp\{\alpha\hat{a}^{\dagger}
-\alpha^{*}\hat{a}\},\;\;
D^{\dagger}(\alpha) = D(-\alpha),\;\; \alpha\in \mathbb{C}.$$
The object looks bleakly abstract to begin with but it is very simple to
visualise. It is a unitary operator depending upon a single complex parameter.
Although the underlying group has no finite dimensional representation, so
that we cannot write it out as a simple matrix, the objects themselves may be
viewed one-to-one as residing on a plane. A picture of this variety is
invaluable throughout, see Fig. \ref{fig1}.
%
%

The commutation relations alone are sufficient to determine the following
group multiplication rule\cite{perel}:
\begin{equation}
\label{multrule}  
D(\alpha)D(\beta) =
\exp\{i{\rm Im}[\alpha\beta^{*}]\} D(\alpha + \beta).
\end{equation}
This simple rule expresses all that we need to know about manipulation of the
abstract operators $D(\alpha)$. It was discovered by Weyl\cite{weyl}, who gave
a simple way to understand it as an encoding of quantal noncommutativity via
the presence of the phase factor ${\rm Im}[\alpha\beta^{*}]$. What we have in
the operators $D(\alpha)$ is a familiar Abelian group, addition on the
complex plane, modified by the inclusion of a leading phase factor. Such a
group representation is termed a {\em projective\/} or {\em ray\/}
representation\cite{weyl}. It would appear that such group representations
must enjoy an intimate relationship with the deep subject of quantal phase
factors\cite{berry1,dirac1a}. 

To understand the importance of the leading phase, let us now consider the
action of the operators $D(\alpha)$ upon some Hilbert space that carries a
representation of them. For example, consider the vacuum ket $|0\rangle$ as
defined by the annihilation condition $\hat{a}|0\rangle = 0$. Of interest
shall be the class of translated vacuum states defined by the rule 
$|\alpha\rangle\equiv D(\alpha)|0\rangle$. As is
well known\cite{klau1,perel,zhang}, these happen to be the familiar minimum
uncertainty states that were first discovered by Schr\"{o}dinger. Under
further translation, the rule (\ref{multrule}) ensures that $|\beta\rangle$
suffers a simple additive translation of its argument $\beta$ to
$\alpha+\beta$:     
\begin{equation}       D(\alpha)e^{i\gamma}|\beta\rangle
=  e^{i(\gamma + \delta\gamma)}   |\alpha+\beta\rangle.     
\end{equation}
Here we have included the phase $\gamma$ to emphasise the nonintegrable
nature of the phase developed by an evolving state $|\alpha\rangle$. It is
related to the quantum geometric phase in a way that we shall later make
explicit. To do this we need to understand its origin. It comes directly from
the rule (\ref{multrule}) as the phase change 
$$\delta\gamma = {\rm Im}[\alpha\beta^{*}]$$ 
Nonintegrability of this phase is clear when we consider a circuit on the
group. The quantity 
$$D(-\alpha)D(-\beta)D(\alpha)D(\beta) =  
\exp\{2i{\rm Im}[\alpha\beta^{*}]\}{\hat I}$$ 
has on the left an expression that describes a circuit on the group manifold,
on the right it has a phase factor that multiplies the unit operator. If we
were to place a ket to the right of both operators then we get a phase that
the ket develops under the circuitous transport dictated by the path on the
group. This is not the Berry phase\cite{berry1}, but we can extract
that from it. Most important of all, the phase change is the same for any
chosen ket.

\subsection{Operator change of variables and the role of $\hbar$}
A result of great practical utility concerns the close correspondence between
the algebra of creation and annihilation operators and that of the canonically
conjugate position and momentum operators. They may be carried one to another
by a mere change of variables. A clear recognition of this property is
important to understanding the role of $\hbar$ within this work.

Suppose we were to take $\hat{p}$ and $\hat{q}$ as primary. Consider now
the new operators
$$\hat{a}  = c_{q}\hat{q}+ic_{p}\hat{p}
\hspace{2cm}
\hat{a}^{\dagger} = c_{q}\hat{q}-ic_{p}\hat{p}.$$
If $\hat{q}$ and $\hat{p}$ satisfy the canonical commutation relation
$[\hat{q},\hat{p}]=i\hbar\hat{I}$ then explicit calculation shows that
$\hat{a}$ and $\hat{a}^{\dagger}$ satisfy $[\hat{a},\hat{a}^{\dagger}] =
\hat{I}$ provided only that we choose the constants $c_{p}$ and $c_{q}$ so
that $c_{q}c_{p}=1/2\hbar$. 

In harmonic oscillator problems one uses the standard change of variables  
$$c_{q} =\sqrt{\omega/2\hbar}  \hspace{2cm} c_{p} =
1/\sqrt{2\hbar\omega},$$  
where $\omega$ has the dimensions of mass by inverse time. If there is no
characteristic time then $\omega$ is free, so it would appear that we cannot
exploit the above connection. 

Surprisingly, this is not true. In Weyl operator calculations one can
generally ignore the constants $c_{q}$ and $c_{p}$. To see why one carries
out the substitutions 
\begin{equation}
\label{change} 
\alpha       = c_{q}q + ic_{p}p \;\;\mbox{and}\;\;
\hat{\alpha} = c_{q}\hat{q} + ic_{p}\hat{p},
\end{equation}
to convert $D(\alpha)$ into its equivalent Weyl form\cite{weylandphase}:
\begin{equation}
\label{weylform}  
U[q,p] \equiv \exp\{i/\hbar(p\hat{q} - q\hat{p})\}.
\end{equation}
It is significant that the $c_{q}$ and $c_{p}$ dependence now appears only
through their product $c_{q}c_{p}=1/2\hbar$. It therefore follows that a free
translation is possible between both pictures by use of (\ref{change}) with
the ratio $c_{q}/c_{p}$ remaining undetermined. So long as due care is
exercised one can set $c_{q}=c_{p}=1$ and re-insert $\hbar$ or either constant,
via the use of dimensionality considerations alone.

\subsection{Quantal Phase Space as an Equivalence Class in $\Psi$}
Let us now construct a natural quantal analogue of classical phase space,
being only a minor variant of that commonly used in the theory of coherent
states\cite{weylandphase}. Rather than use the Weyl operator labels $(q,p)$
as the analogue of classical phase space, we prefer to construct phase space
as an equivalence class $\tilde{\psi}(q,p)$ of states $\psi$ that share the
same expectation values for $\hat{p}$ and $\hat{q}$. 

First let us review properties of the Weyl group action upon arbitrary kets.
Consider the Weyl translates of an arbitrary state labelled $|\phi\rangle$. 
Of interest therefore is the class of states:
$$|q,p;\phi\rangle = U[q,p] |\phi\rangle.$$ 
As is well known, the linear action of $U[q,p]$ effects a mere additive
translation of the expectation values of any given state $\phi$. The reason
for this may be traced to the simple relationships\cite{klau1}: 
\begin{equation}
U[q,p]^{\dagger} \hat{p} U[q,p] = \hat{p} + p\hat{I}
\;\; \mbox{and} \;\;
U[q,p]^{\dagger} \hat{q} U[q,p] = \hat{q} + q\hat{I}.
\end{equation}
As a consequence of this fact it is then clear that given any $\phi$, one can
compute its position and momentum expectation values and then use these
parameters in a Weyl operator so as to translate $\phi$ back to a special
representative state  
$\phi_{0}\equiv 
U[-\langle \hat{q}\rangle_{\phi}, -\langle \hat{p}\rangle_{\phi}]
|\phi\rangle$ characterised by the property that both expectation values
are equal to zero. In this manner one obtains a convenient parametrisation
of the entire Hilbert space ${\cal H}$ in terms of the Weyl translates 
$U[q,p] |\phi_{0}\rangle$ of all states $\phi_{0}$ having both expectation
values equal to zero. This result is of some utility in picturing the entire
Hilbert space as being generated out of all states $\phi_{0}$, see Fig.
\ref{fig2}.
%
%

Of related utility is the following {\em coordinate\/} map
\begin{equation}
\label{equivdef}   
\Pi:{\cal H}\mapsto \mathbb{R}^{2}  \;\;\mbox{where}\;\;
\Pi[\psi] = (q(\psi),p(\psi));\;\;  q(\psi) \equiv\langle
\hat{q}\rangle_{\psi}\;\;\mbox{and}\;\; p(\psi) \equiv\langle
\hat{p}\rangle_{\psi}. \end{equation} 
This rule throws away the $\psi$--dependence and enables us to extract
classical coordinates. It functions here as a mathematical device for
defining a quantal analogue of classical phase space as the natural
$\Pi$-induced equivalence class:  
\begin{equation}
\label{equivclass} 
\tilde{\psi}(q,p) =\{\psi\in{\cal H}: 
\Pi[\psi] = (q,p)\in \mathbb{R}^{2}\}.
\end{equation}
The interpretation of this rule is simple. We partition ${\cal H}$ into all
possible sets of states $\tilde{\psi}(q,p)$ which happen to share the
same expectation values for position and momentum operators. A particular
class $\tilde{\psi}(q,p)$ can now be generated via application of $U[q,p]$ to
the special class $\tilde{\psi}(0,0)$ consisting of all $\phi_{0}$ such that
$\Pi[\phi_{0}] = (0,0)$. This class of states plays the role of an origin
in the quantal phase space so constructed\cite{origin}.

To make the above ideas concrete, we consider now the explicit nature of
$\tilde{\psi}(q,p)$ in the coordinate representation of a one degree of
freedom system. To avoid confusion between operators and the group parameters
it is helpful to first relabel $q$ and $p$ as $Q$ and $P$. Then one can use
the definition (\ref{weylform}), and an elementary application of the
Baker--Campbell--Hausdorff formula\cite{perel}, 
$e^A e^B = e^{1/2[A,B]}e^{A+B}$, for $[A,[A,B]]=0$ and $[B,[A,B]]=0$, so as to
disentangle the action of $\hat{q}$ and $\hat{p}$. If at the same time one
makes passage to the Schr\"{o}dinger representation, via the substitutions
$\hat{q}\mapsto q$ and $\hat{p}\mapsto -i\hbar \partial/\partial q$, then the
relabelled operators $U[Q,P]$ can be written in either of the two convenient
reordered forms:  
\begin{eqnarray}
\label{wellord}   
U[Q,P] & = & \exp\{+iPQ/2\hbar\} 
\exp\{-Q\partial/\partial q\}
\exp\{iPq/\hbar\},\\
\label{antiord} 
U[Q,P] & = & \exp \{-iPQ/2\hbar\}
\exp\{iPq/\hbar\}
\exp\{-Q\partial/\partial q\}.
\end{eqnarray}
Use of the operational calculus identity
$$\exp\{-Q\partial/\partial q\}f(q) = f(q-Q),$$ 
then enables one to compute, via either route, the general result\cite{perel}
\begin{equation}
U[Q,P]\phi(q) = \exp \{-iPQ/2\hbar\} \exp\{iPq/\hbar\}
\phi(q - Q) \in \tilde{\psi}(q,p).
\end{equation}
If one now lets $\phi_{0}(q)$ be an arbitrary member of the class
$\tilde{\psi}(0,0)$, then it is clear that $\psi(q) \equiv e^{-iPQ/2\hbar}
e^{iPq/\hbar} \phi_{0}(q - Q)$ is an arbitrary member of the class
$\tilde{\psi}(Q,P)$. 

\section{Weyl operators as propagators}
\label{intuit}
Putting together the preceding material, we are led to the following rather
striking observation. If one considers an arbitrary classical Hamiltonian
system $H_{c}(q,p)$ and we pick a particular solution, say  $(q(t),p(t))$,
then it is a pure triviality to notice that the lifting of this
trajectory into the argument of a time--dependent Weyl operator $U[q(t),p(t)]$
creates a propagator which reproduces the classical trajectory as a
time--dependent equivalence class of states $\tilde{\psi}(q(t),p(t))$.
Moreover, it is a straightforward property of Weyl translation that the
resulting evolution  must be non--dispersive for all member states of the 
time--dependent class $\tilde{\psi}(q(t),p(t))$. 

Therefore, we are led to the non--trivial conclusion that if one can find a
way to generate a dynamics of Weyl operators, such that the arguments
of these follow the classical trajectories of any chosen classical system,
then one will have a wavefunction dynamics for the equivalence classes
$\tilde{\psi}(q,p)$ such that their expectation values also follow the
desired classical trajectories. Within such a scheme, one can then see how 
the macroscopically successful theory of classical mechanics might well just
amount to a special form of quantum mechanics where the {\em equivalence 
class\/} is known but not the  actual {\em wavefunction\/}. 

The remarkable adjunct to this conclusion is the knowledge that the Weyl
operator action within ordinary linear quantum mechanics is trivial, it is
just linear propagation upon the plane. But that very triviality aids us in a
totally surprising way. Because of the linear group action, we cannot
achieve the desired goal by using a linear theory of quantum mechanics. But
the companion of that statement is that because of this very same linear
action, there does exist a natural nonlinear theory which does it.

The trick is that any smooth trajectory is built up out of infinitesimal
linear segments. As the generator of a given infinitesimal Weyl propagator
$U[\delta q,\delta p]$, that adds increments $\delta q$ and $\delta p$ to the
coordinates of some $\tilde{\psi}(q,p)$, one might consider the possibility
of taking a quantal {\em phase space dependent\/} Hamiltonian operator
$\hat{H}(\tilde{\psi})$. If one were to allow this operator to vary across
the quantal phase plane then it is intuitively plausible to suppose that one
might build up the right dynamics precisely because of what seemed to
be a problematic linear action of the Heisenberg--Weyl group.
 
\section{A direct geometric connection}
To pursue this idea we now need a way to get the required quantal Hamiltonian
from knowledge of the classical one. Astonishingly, one can do this in a very
simple manner by exploiting the geometrical content of classical mechanics. 

Recall that the Hamiltonian formulation of classical dynamics defines a vector
field for the flow of phase space points according to the formul{\ae}:    
\begin{equation} \label{eq:class}
\dot p  =  -\frac{\partial H_{c}}{\partial q}\;\;\mbox{and}\;\;
\dot q  =  \frac{\partial H_{c}}{\partial p},
\end{equation} 
where of course $H_{c}(q,p)$ is the classical Hamiltonian. Given a family of
trajectories in classical phase space, let us pick one, say $(q(t),p(t))$.
Having chosen to parametrise this curve with time differentiation of
it must yield the local dynamical field $(\partial H_{c}/\partial p,-\partial
H_{c}/\partial q)$ as a consequence of (\ref{eq:class}). A similar property 
is shared by the evolution operators of quantum mechanics. Given $U(t,t_{0})$
one can deduce $\hat{H}(t)$ by differentiation and use of the simple rule
$\hat{H}(t)= i\hbar[d/dt U(t,t_{0})]U(t,t_{0})^{-1}$. This trick provides
the essential geometric connection that we shall need to forge a natural
nonlinear quantization condition which provides the required
$\hat{H}(\tilde{\psi})$ in a manner that is unique up to a multiple of the
unit operator. The latter freedom shall subsequently be removed via energetic
considerations.

\section{A ``Classical'' quantization procedure}
Recall that the motion of classical phase space points is to be implemented
using elements of the equivalence classes $\tilde{\psi}(q,p)$. To transport
these in a manner that duplicates the classical motion it is clear that we
must lay down upon the plane of parameters of the Weyl operators an 
operator--valued analogue of the Hamiltonian vector field pertaining to 
the classical system of interest. 

Starting with a family of classical trajectories that are the solutions of a
particular classical Hamiltonian $H_{c}(q,p)$, we shall simply project these
onto the quantal phase space so as to obtain the required quantal
operator--valued field. Throughout, it helps to keep in mind two copies of
the phase plane. In the classical plane dynamics is the motion of phase space
points; in the quantal plane dynamics is the motion of wavefunctions or
operators that happen to be indexed by complex numbers. To underline this we
shall initially reserve explicitly complex quantities as referring to the
quantal phase plane and explicitly real quantities as referring to the
classical phase plane. The situation is depicted in Fig. \ref{fig3}.
%
%
 
The required projection is the standard association
$$c_{q}q(t) + ic_{p} p(t) \mapsto \alpha(t),$$
where the parameter $t$ is time, but $H_{c}(q,p)$ need not be explicitly 
time--dependent and the constants $c_{q}$ and $c_{p}$ are free. 

Working with Weyl operators in the $D(\alpha)$ form we start by seeking the
unique quantal Hamiltonian that generates the following desired  group
trajectory:     
$$\tilde{D}(\alpha(t)) \equiv \exp
\left\{i\int_{0}^{t}{\rm Im}[\alpha^{*}(\tau)\dot{\alpha}(\tau)]
\,d\tau\right\}  D(\alpha(t)),$$  
where $\alpha(t) = c_{q}q(t) + ic_{p}p(t)$ and the leading phase has been
chosen purely for convenience, but shall later be fixed via energetic
considerations. 

In virtue of the linear action of the operators $D(\alpha)$ one can easily
guess that the generator of such a trajectory must be the infinitesimal Weyl
operator $D(\dot{\alpha}\delta t)$. This guess is supported by using the rule
(\ref{multrule}) to evaluate the product $D(\dot{\alpha}\delta t)D(\alpha(t))$
so as to find the expected result
\begin{equation}
\label{guess}
D(\dot{\alpha}\delta t)D(\alpha(t))=
\exp\{i{\rm Im}[\alpha^{*}\dot{\alpha}]\delta t\} \times D(\alpha(t) +
\dot{\alpha}\delta t). 
\end{equation}
To prove that this guess is correct we need only differentiate the trajectory
of the Weyl operator $\tilde{D}(\alpha(t))$. From appendix one we obtain the
identity
\begin{equation}
\label{derivative}  
\frac{d}{dt}\tilde{D}(\alpha(t)) = 
(\dot{\alpha}\hat{a}^{\dagger} - \dot{\alpha}^{*}\hat{a})
\tilde{D}(\alpha(t)), 
\end{equation}
so that we may now read off the required quantal Hamiltonian. Doing that we
discover that $\hat{H}_{q}(\alpha)$ is an implicit function of $\alpha$ given
at this stage only in terms of $\dot{\alpha}$ by the expression:  
\begin{equation}
\label{hamdef}
\hat{H}_{q}(\alpha) =
i(\dot{\alpha}\hat{a}^{\dagger} - \dot{\alpha}^{*}\hat{a}),
\end{equation}
To remove the implicit nature of this connection, we now use the 
correspondence constraint, $\alpha = c_{q}q+ ic_{p}p$, its derivative
$\dot{\alpha} = c_{q}\dot{q} + ic_{p}\dot{p}$, and the fact that the validity
of Hamilton's equations for the original classical system ensures that 
$$(\dot{q},\dot{p}) = 
(\partial H_{c}/\partial p,-\partial H_{c}/\partial q).$$
Solving for $\dot{\alpha}$ one discovers the quantization condition:
\begin{equation}
\label{alphadot}   
\hat{H}_{q}(\alpha) =  
i\left(c_{q}\frac{\partial H_{c}}{\partial p} - 
ic_{p}\frac{\partial H_{c}}{\partial q}\right)\hat{a}^{\dagger}
+
\left(c_{q}\frac{\partial H_{c}}{\partial p} + 
ic_{p}\frac{\partial H_{c}}{\partial q}\right)\hat{a},
\end{equation}
where it should be stressed that the the derivatives of the classical
Hamiltonian are to be evaluated upon the quantal phase plane at the point
specified by the rule $(q,p) \leftrightarrow \alpha= c_{q}q+ ic_{p}p$, where
$\alpha^{*}= c_{q}q- ic_{p}p$.

The relation (\ref{alphadot}) thus fixes $\hat{H}_{q}(\alpha)$ in terms of the
projected derivatives of the classical Hamiltonian from the classical phase
plane to the quantal phase plane, at the corresponding point and at the
corresponding time. Our quantization condition is thus uniquely derived, 
modulo questions of phase arbitrariness, by the purely geometric constraint
that the Weyl operator dynamics should reproduce the classical dynamics for
{\em all possible\/} $H_{c}(q,p)$. 

It is helpful to now drop the convention that quantal phase plane quantities
should be complex. The algebraic changes: $\alpha= c_{q}q+ ic_{p}p$ and
$\hat{a} =c_{q}\hat{q}+ ic_{p}{p}$ with $c_{q}c_{p}=1/2\hbar$, now convert
$\tilde{D}(\alpha(t))$ into its equivalent form
\begin{equation}
\label{propagator}
\tilde{U}[q(t),p(t)] = 
\exp\left\{\frac{-i}{2\hbar}\int_{0}^{t}
(p(\tau)\dot{q}(\tau) - q(\tau)\dot{p}(\tau))\,d\tau\right\}
U[q(t),p(t)],
\end{equation}
while equations (\ref{derivative}) and (\ref{alphadot}) become:
\begin{eqnarray}
\label{evolve}
\frac{d}{dt}\tilde{U}[q(t),p(t)] & = &
\frac{-i}{\hbar}(\dot{q}\hat{p} - \dot{p}\hat{q})
\tilde{U}[q(t),p(t)],\\
\label{hamdef2} 
\hat{H}_{q}(q,p) & = &
\frac{\partial H_{c}}{\partial q}(q,p)\hat{q} +
\frac{\partial H_{c}}{\partial p}(q,p)\hat{p}.
\end{eqnarray}
In this manner the choice of scale in both the original classical system and
that of its formal quantal correspondent is made identical. Moreover, it is
readily seen that a change in $\hbar$ has no effect other than that of
rescaling all classical trajectories.

\section{Discussion of the result}
The existence of this remarkable dynamical correspondence depends upon the
highly nontrivial fact that the Weyl group action upon the phase plane is
linear. Because of this fact the ``tangent'' vectors to the trajectories in
both classical mechanics and this special Weyl operator dynamics are just the
derivatives of the phase plane parameters. Since the two parameter spaces are
isomorphic, and the depth of Hamilton's equations is just the statement that
$H_{c}(q,p)$ specifies how the field of tangent vectors should vary across
the phase plane, we are able to achieve a direct quantization that generates
an isomorphic dynamics of Weyl operators. 

For those concerned about our use of the time $t$, let it be noted that the
geometric nature of our argument ensures that $t$ appears in this derivation
purely as a parameter that enforces a synchronicity between two
identical families of trajectories. It ties the temporal development of
individual points in the classical and quantal planes one-to-another. That is
why we are able to both use it and then later remove it. Therefore, it should
be stressed that there are no assumptions of adiabaticity, or anything of
that nature. If the initial classical Hamiltonian were to contain an explicit
time dependence then the right hand side of (\ref{alphadot}) would also be
explicitly time--dependent and would so fix an operator valued field that
varies over the group in time. 

A direct view of the result is offered upon consideration of the small time
propagator argument that motivated it. Using (\ref{hamdef2}) in the small
time Weyl propagator:
$$U_{\delta t}[q,p]
\equiv \exp\left\{\frac{-i}{\hbar}\left[
\frac{\partial H_{c}}{\partial q}\hat{q} +
\frac{\partial H_{c}}{\partial p}\hat{p}
\right]\delta{t}\right\},$$
one computes the new group element
$$U_{\delta t}[q,p]\cdot\tilde{U}[q,p] =
\exp\left\{\frac{-i}{2\hbar}
\left[p \frac{\partial H_{c}}{\partial p} + 
      q \frac{\partial H_{c}}{\partial q}\right]\delta t\right\}
U[q + \frac{\partial H_{c}}{\partial p}\delta t,
  p - \frac{\partial H_{c}}{\partial q}\delta t]$$
which must be $\tilde{U}[q(t+\delta t),p(t+\delta t)]$. So having carried the
classical dynamical field down to the quantal phase plane, one sees directly
that the integration of the equation of motion for our propagator corresponds
precisely to an integration of the original classical equations in Hamiltonian form.

Note that the well known quantum--classical dynamical
correspondence\cite{zhang} for Hamiltonians which are quadratic or
less in $\hat{q}$ and $\hat{p}$ is separate from our result. However, it is
worthy of note that an early paper by Klauder\cite{klau2}, upon the action
formulation of quantum dynamics, passes very close by the discovery made here
that a complete quantum to classical dynamical correspondence is possible once we permit a nonlinear Schr\"{o}dinger equation involving a 
state--dependent Hamiltonian. There it was shown that one can reproduce the 
dynamics of the Harmonic oscillator using only a linear combination of the
operators $\hat{q}$ and $\hat{p}$. 

On a deeper level, our entire result depends upon the remarkable mathematical
fortuity that the Heisenberg--Weyl group affords a ray representation of the
Abelian group of translations on the complex plane\cite{weyl}. As such, we
have uncovered something which is mathematically obvious, but perhaps
physically unexpected.  

\clearpage
\section{A ``Classical'' Schr\"{o}dinger equation}
Thus far we have been concerned with dynamics upon the Heisenberg--Weyl group.
This must now be carried to the quantum states. 

Combining equations (\ref{evolve}) and (\ref{hamdef2}) we obtain the general
Heisenberg--Weyl group evolution equation
\begin{equation}
i\hbar\frac{d}{dt}\tilde{U}[q,p] =
\hat{H}_{q}(q,p)
\tilde{U}[q,p];\;\;\mbox{where}\;\;
\hat{H}_{q}(q,p) = 
\frac{\partial H_{c}}{\partial q}(q,p)\hat{q} +
\frac{\partial H_{c}}{\partial p}(q,p)\hat{p}
\end{equation}
where the coordinates have meaning with respect to the group manifold. Given
this abstract dynamics, it is clear that we can place an arbitrary quantum
state $\psi$ to the right, and so obtain an evolution equation for quantum
states. The problem is then how to fix $(q,p)$ from knowledge of $\psi$.

Since $\tilde{U}[q(t),p(t)]$ simply translates expectation values, it is clear
that one must carry the abstract Weyl dynamics to the quantum states $\psi$
by fixing $(q,p)$ in the operator $\hat{H}_{q}(q,p)$ in terms of the
coordinates  
\begin{equation}
\label{coords} 
q(\psi) = \langle\hat{q}\rangle_{\psi}\;\;\mbox{and}\;\;
p(\psi) = \langle\hat{p}\rangle_{\psi}
\end{equation}
of the equivalence class $\tilde{\psi}(q,p)$ to which $\psi$ belongs.

Doing that we arrive at the formal nonlinear Schr\"{o}dinger equation
\begin{equation}
\label{eq:quan}
i\hbar\frac{d}{dt}|\psi\rangle
= \hat{H}_{q}(\tilde{\psi})|\psi\rangle.
\end{equation}
It is important to understand that this prescription is arrived at in order
that our dynamical correspondence construction should work. There is no
empirical evidence for its physical validity but the possibility appears
worthy of some attention.

\subsection{A generalized dynamical law}
In this respect, it is remarkable that a dynamics defined by (\ref{eq:quan})
is always norm--preserving provided only that 
$\hat{H}_{q}(\tilde{\psi}) =
\hat{H}_{q}^{\dagger}(\tilde{\psi}^{*}).$
To prove this one uses (\ref{eq:quan}) and its conjugate to find that
$$i\hbar\frac{d}{dt}\langle \psi|\psi\rangle
 =  \langle \dot{\psi}|\psi\rangle +
      \langle\psi|\dot{\psi}\rangle
 = 
\langle \psi| 
\left[-\hat{H}_{q}^{\dagger}(\tilde{\psi^{*}})
      +\hat{H}_{q}(\tilde{\psi})\right]
|\psi\rangle \\
 =  0.$$
Thus $||\psi||^{2}$ remains an invariant. However, unlike linear quantum
theory the Hilbert space inner product $\langle \phi|\psi\rangle$ is not.
The reason for this lies in a repeat of the above argument, where one is
led to discover that
$$i\hbar\frac{d}{dt}\langle \phi|\psi\rangle
= \langle \phi| 
\left[-\hat{H}_{q}^{\dagger}(\tilde{\phi^{*}})
      +\hat{H}_{q}(\tilde{\psi})\right]
|\psi\rangle\ne 0.$$
In understanding this fact it is instructive to note that the presence of the
$\tilde{\psi}$--dependence is formally equivalent to the case of a 
time--dependent Hamiltonian, albeit one such that different states $\phi$ and $\psi$ evolve under different Hamiltonians. 

The loss of an invariant inner product means that we have now lost much of
the Hilbert space structure of linear quantum mechanics. The dynamical arena
is now the Banach space\cite{banach} of all square integrable $\psi$, since we
retain an invariant norm $||\psi||^{2}$. This loss of structure does not
however mean that the usual Hilbert space methods are useless. One can still
use the device of referring the evolution of a given state $\psi$ with respect
to a static orthonormal basis in an auxilliary space\cite{klau3}. 

Given a countable orthonormal basis $\{\phi_{k}\}_{k=1}^{\infty}$ one can
write any $\psi$ in terms of the coordinate functionals $C_{k}[\psi(t)]
\equiv \langle \phi_{k}|\psi(t)\rangle.$  The intuitive meaning of a
nonlinear operator on such a space is then that of an infinite number of
nonlinear relations $C'_{k}(C_{1},C_{2},\ldots)$ among the initial
coordinates $C_{k}$ and the final coordinates $C'_{k}$. Knowledge of the
$C'_{k}$ then enables reconstruction of the transformed state $\psi'$.
However, it is clear that such an idea will not work in general because
convergence of the new norm ${C'}_{k}{C'}_{k}^{*}$ is not assured. Here we 
are fortunate in that the bootstrap from linear Hermitian generators in the
construction of (\ref{eq:quan}) assures us that this will be so. This simple
fact means that one might legitimately expect (\ref{eq:quan}) to have a 
rich mathematical theory. Having said that, difficult problems remain in the
elucidation of conditions under which the $\psi$--dependence must enter into
$\hat{H}_{q}(\tilde{\psi})$. 

\subsection{Minimal invariance}
As a temporary measure, it seems appropriate to introduce a concept we
call {\em minimal invariance\/}. This means invariance of the equations of
motion with respect to the ordinary globally linear unitary transformations
of quantum mechanics. The idea is that (\ref{eq:quan}) should, at the very
least, share the representation independence of linear quantum theory. 

Consider therefore the action of a linear operator $U$ in the equation
(\ref{eq:quan}). The suggested transformation law is:
\begin{equation}
\label{transform}
|\psi\rangle \mapsto U|\psi\rangle\;\;\mbox{and}\;\;
\hat{H}_{q}(\tilde{\psi})\mapsto
U\hat{H}_{q}(U\tilde{\psi})U^{\dagger},
\end{equation}
where the essential new feature occurs in the carrying of $U$ to the argument
of $\hat{H}_{q}(\tilde{\psi})$. 

Examination of this condition, and the Hermiticity constraint
$\hat{H}_{q}(\tilde{\psi}) = \hat{H}_{q}^{\dagger}(\tilde{\psi}^{*})$, is
then suggestive of the need to have $\psi$ enter into
$\hat{H}_{q}(\tilde{\psi})$ purely via quantities that involve $\psi$ and
$\psi^{*}$ in a real-valued and representation independent combination. 
Expectation values, such as the coordinates (\ref{coords}) of our model
nonlinear system, have this property so that minimal invariance is assured in
this case. For example, application of (\ref{transform}) to the quantization
rule (\ref{hamdef2}) leaves the arguments of the coefficients $\partial
H_{c}/\partial q$ and $\partial H_{c}/\partial p$ invariant, and hence their
values, while taking $\hat{q}$ and $\hat{p}$ into the appropriately
transformed operators $U\hat{q}U^{\dagger}$ and $U\hat{p}U^{\dagger}$. 

As a finite dimensional example of the same idea, one might consider an
ordinary linear Hamiltonian, with the spectral resolution:
$$\hat{H} = \sum_{k=1}^{n} E_{k} |\phi_{k}\rangle\langle \phi_{k}|.$$
This can be made nonlinear in the sense of (\ref{eq:quan}) by the device
of multiplying the eigenvalues by representation independent quantities.
An interesting example of such a system is the model:
$$\hat{H}(\psi) \equiv
\sum_{k=1}^{n} E_{k}(|\langle\phi_{k}| \psi\rangle|^{2})\cdot
|\phi_{k}\rangle\langle \phi_{k}|.$$
One has on the space of states $\psi\in \mathbb{CP}^{n-1}$ a smoothly varying
operator field  $\hat{H}(\psi)$, such that the original stationary states
$\phi_{k}$ remain, as do the energy eigenvalues $E_{k}$. However, away from
these unperturbed stationary states the dynamics is quite different. It
remains completely integrable, but now includes shear motion.

\subsection{A possible physical interpretation}
One formal way to view the dynamical law (\ref{eq:quan}) is as the natural
generalization of linear quantum mechanics under the postulate that {\em
global unitarity\/} in function space should be relaxed into the wider system
of {\em local unitarity\/}, where the intended concept of locality refers to
points in the abstract topological space of all possible $\psi$. There is
no physical motivation for that view other than as an exploratory route for
the purpose of generalization.

A more physical approach would be to interpret a state--dependent Hamiltonian
as standing for elements of back-reaction of a system upon its environment. If
$\psi$ is all, if $\psi$ is a real entity; then it ought to be the case that
the value of $\psi$ for a large and non--isolated system should determine in
some appropriate sense its future evolution and hence its Hamiltonian. An 
approach of this kind is reminiscent of Everett's Universal $\Psi$\cite{eve}.
On another tack, it is considered most encouraging that it is precisely
within the classical regime, that a dynamics of the type (\ref{eq:quan})
finds correspondence with a successful physical theory. 

However, it appears premature to devote too much attention to the general
rule (\ref{eq:quan}). Many questions of principle remain to be solved. The
rest of this paper is therefore concerned with refining understanding of the
particular system defined by (\ref{hamdef2}) rather than general consideration
of (\ref{eq:quan}). As a final remark on that topic it is interesting that
the general theory of Weinberg\cite{wein1,wein2} shares the property of
norm--preserving dynamics. It seems that a completely general theory based
upon (\ref{eq:quan}) would have to be equivalent to that of
Weinberg\cite{equiv}.

\section{Correcting the Energy}
Since our geometrical construction retains expectation values for the
coordinates, we make the postulate that these remain the correct way to
extract the values of general state--dependent observables. To the operator
$\hat{A}({\tilde{\psi}})$ we then associate the observable value:
\begin{equation}
\label{expect}
A(\tilde{\psi}) = \langle\psi|\hat{A}(\tilde{\psi})|\psi\rangle.
\end{equation}
It is remarkable that this postulate works, most especially so in respect
of the well-known observation that the standard probabilistic interpretation
of $\psi$ cannot hold in the absence of an invariant Hilbert space inner
product\cite{jordan}. Presently, it is not clear to us that this fact should
rightly be considered a {\em problem\/} or perhaps a {\em potential solution to a problem\/}; particularly so in connection with the deep unresolved problems of quantum measurement theory\cite{qmt,pearle}. 

Our postulated rule (\ref{expect}) therefore has merit in so far as it is
successful for the purpose of constructing a quantal model of classical
mechanics. Moreover, it appears necessary in order to achieve consistency with
our use of the coordinate rule (\ref{coords}). 

Using (\ref{expect}) and the operator (\ref{hamdef2}) we can now compute
the energy of our model classical system when in the state $\psi$: 
\begin{equation} 
\label{e0}
E_{0}(\psi) = 
\frac{\partial H_{c}}
{\partial q}(q,p)\cdot
\langle\hat{q}\rangle_{\psi} + 
\frac{\partial H_{c}}
{\partial p}(q,p)\cdot
\langle\hat{p}\rangle_{\psi};\;\;
(q,p)= (\langle\hat{q}\rangle_{\psi},\langle\hat{p}\rangle_{\psi}).
\end{equation}
This is certainly not the desired exact classical energy. 

However, a most interesting possibility now presents itself. If one recalls
that there was an essential time--dependent phase arbitrariness in the
mapping to Weyl operator trajectories, then there appears an avenue towards
the dual resolution of two problems. Let us therefore explore the idea that a
removal of the quantal phase freedom might be connected with the necessity of
correcting the classical energy!

To pursue an energetic correspondence with classical mechanics we now invoke
the remaining freedom to add a multiple of the unit operator to the
Hamiltonian $H(\tilde{\psi})$ as the sole available means of altering the
value of our energy observable, $E(\tilde{\psi})$, without affecting the
desired classical to quantal dynamical correspondence.

Since both the desired energy,
$E(\tilde{\psi})
= H_{c}(\langle\hat{q}\rangle_{\psi},\langle\hat{p}\rangle_{\psi})$,
and the original energy (\ref{e0}) are pure numbers, we need only
take their difference so as to construct the correction operator:
\begin{equation}
\left[H_{c}
- 
\frac{\partial H_{c}}
{\partial q}\cdot
\langle\hat{q}\rangle_{\psi} 
-
\frac{\partial H_{c}}
{\partial p}\cdot
\langle\hat{p}\rangle_{\psi}\right]\hat{I},
\end{equation}
as being that unique multiple of the unit operator which we must add to the
original quantization condition (\ref{hamdef2}). 

The residual phase arbitrariness of the original derivation is thereby removed
and we arrive at a uniquely determined quantization condition: 
\begin{equation} 
\label{nlham2} 
\hat{H}_{q}(\tilde{\psi})\equiv
H_{c}(\tilde{\psi}) + \frac{\partial H_{c}}{\partial p}(\tilde{\psi}) 
(\hat{p} -\langle\hat{p}\rangle_{\psi})   
+ \frac{\partial H_{c}}{\partial q}(\tilde{\psi}) 
(\hat{q} - \langle\hat{q}\rangle_{\psi}).
\end{equation}
In pausing to consider this situation, we find it quite remarkable that the
entire procedure is successful to this degree of correspondence. Not only
have we recovered exact classical dynamics, but also the correct energy.
Moreover all of this is achieved with a seemingly perfect blend between both
classical and quantal concepts. Although some scepticism about the postulates
made appears warranted, we find it truly incredible that a construction
having this level of completeness is even possible. 

\section{Ehrenfest's Theorem}
As a prelude to discussion of Poisson brackets, commutators and the inclusion
of classical canonical transformations, it is instructive to verify that the
prescription (\ref{nlham2}) actually does recover classical dynamics in
Hamiltonian form. The discussion exposes a rather special property of our
quantal phase space of equivalence classes.

We begin by making an assumption about the quantal phase space coordinates
$\langle \hat{q} \rangle_{\psi}$ and $\langle \hat{p} \rangle_{\psi}$. Let
us suppose initially that the operators $\hat{q}$ and $\hat{p}$ do
not depend upon $\psi$. As a result of this assumption it follows that:
\begin{eqnarray}
\frac{d}{dt} \langle \hat{q} \rangle_{\psi}
& = & \langle \dot{\psi}|\hat{q}|\psi\rangle + 
      \langle \psi|\hat{q}|\dot{\psi}\rangle,\\
\frac{d}{dt} \langle \hat{p} \rangle_{\psi}
& = & \langle \dot{\psi}|\hat{p}|\psi\rangle + 
      \langle \psi|\hat{p}|\dot{\psi}\rangle.
\end{eqnarray}
Using this decomposition, we now invoke the defining equation of motion
(\ref{eq:quan}) and the quantization rule (\ref{nlham2}) to discover that
\begin{eqnarray}
\label{eh1}
\frac{d}{dt} \langle \hat{q} \rangle_{\psi}
& = & \frac{1}{i\hbar} \left\langle
[\hat{q},\hat{H}_{q}(\tilde{\psi})]\right\rangle 
= \frac{\partial H_{c}}{\partial p}
(\langle \hat{q} \rangle_{\psi},\langle \hat{p} \rangle_{\psi}),\\
\label{eh2}
\frac{d}{dt} \langle \hat{p} \rangle_{\psi}
& = & \frac{1}{i\hbar} \left\langle
[\hat{p},\hat{H}_{q}(\tilde{\psi})]\right\rangle 
= -\frac{\partial H_{c}}{\partial q}
(\langle \hat{q} \rangle_{\psi},\langle \hat{p} \rangle_{\psi}).
\end{eqnarray}
Thus we recover Hamilton's equations, and a special version of Ehrenfest's
theorem\cite{gott} such that the expectation value angle brackets now
lie within the argument of the Hamiltonian derivatives, rather than
surrounding them. 

\subsection{The standard classical limit}
Standard proofs of the classical limit\cite{gott} involve an argument designed
to recover the above couplet of equations in the limit $\hbar\rightarrow 0$,
while making assumptions about the states $\psi$ and the behaviour of
$\partial H_{c}/\partial q$. With that focus it seems that people either
neglected to notice, or did not deem it important, that the above desired
limit holds irrespective of the value of $\hbar$. It thus permits an exact
reproduction of classical mechanics for a fixed non--zero value of $\hbar$.
The importance this work is that we have now provided a natural route for
deduction of the appropriate quantization prescription (\ref{nlham2}), where
the chosen route {\em does not require\/} us to know the ordinary quantization
prescription of linear quantum theory. 

In this connection, it should be noted that our examination of the literature
has uncovered a recent reformulation of linear quantum theory due to
Kay\cite{kay}, wherein the appropriate quantization condition (\ref{nlham2})
appears in the course of an argument he uses to establish the standard
classical limit. A similar result is implicit in Messiah's classic
book\cite{mess}. However, in both cases one starts from knowledge of the
ordinary quantization prescription and the connection with nonlinear quantum
theory is not recognised. 

Later we shall show that continuation of the rule (\ref{nlham2}) to all orders
in the implicit Taylor series enables recovery of the usual quantization
prescription of linear quantum theory. It then follows that the classical
limit may now be viewed in a completely $\hbar$--independent fashion
as the truncation of an operator Taylor series at the first--order term.
In that view appropriate to the current orthodoxy, classical mechanics thus
appears as a dynamically {\em nonlinear\/} approximation to {\em linear\/}
quantum theory.

\subsection{The case of nonlinear coordinate observables}
Our preceding derivation was predicated upon the assumption of coordinate
observables derived from linear operators. As a primer for discussion of
Poisson brackets it is therefore useful to consider the more general case of
coordinate observables defined by the implicit relations: 
$$\langle Q_{c}
\rangle =  Q_{c}(\langle \hat{q} \rangle_{\psi},\langle \hat{p}
\rangle_{\psi}) \;\;\mbox{and}\;\; 
\langle P_{c} \rangle =  P_{c}(\langle
\hat{q} \rangle_{\psi},\langle \hat{p} \rangle_{\psi}).$$ 
This choice is motivated by consideration of the freedom we have to make an
arbitrary canonical transformation in the original classical system after
its quantization via the rule (\ref{nlham2}). Since we deal with a finite
canonical transformation, $Q_{c}(q,p)$ and $P_{c}(q,p)$ are not arbitrary. 
To preserve the Poisson structure they must satisfy the essential condition
\begin{equation}
\label{classcanon}
\{Q_{c},P_{c}\} = \frac{\partial Q_{c}}{\partial q}
\frac{\partial P_{c}}{\partial p} -
\frac{\partial Q_{c}}{\partial p}
\frac{\partial P_{c}}{\partial q}= 1,
\end{equation}
where $\{\bullet,\bullet\}$ denotes the usual Poisson bracket of classical
mechanics. 

Considering the evolution of $\langle Q_{c} \rangle$ and 
$\langle P_{c} \rangle$ one now finds that:
\begin{eqnarray}
\frac{d}{dt}\langle Q_{c} \rangle
& = & \frac{\partial Q_{c}}{\partial q}\cdot 
\frac{d}{dt}\langle \hat{q} \rangle_{\psi} +
\frac{\partial Q_{c}}{\partial p}\cdot 
\frac{d}{dt}\langle \hat{p} \rangle_{\psi}\\
\frac{d}{dt}\langle P_{c} \rangle
& = & \frac{\partial P_{c}}{\partial q}\cdot 
\frac{d}{dt}\langle \hat{q} \rangle_{\psi} +
\frac{\partial P_{c}}{\partial p}\cdot 
\frac{d}{dt}\langle \hat{p} \rangle_{\psi}
\end{eqnarray}
It is now possible to combine this with equations (\ref{eh1}) and (\ref{eh2})
so as to compute the new relationships:
\begin{eqnarray}
\frac{d}{dt}\langle Q_{c} \rangle
& = & \langle\{Q_{c},H_{c}\}\rangle\\
\frac{d}{dt}\langle P_{c} \rangle
& = & \langle\{P_{c},H_{c}\}\rangle
\end{eqnarray}
where the appearance of the classical Poisson bracket is most significant.

In general, the operators $\hat{Q}$ and $\hat{P}$ associated with 
$\langle Q_{c} \rangle$ and $\langle P_{c} \rangle$ must now depend upon
$\tilde{\psi}$. However, working backwards, it becomes clear they can be given
explicitly in terms of the linear operators $\hat{q}$ and $\hat{p}$, via
the rule (\ref{nlham2}), as the two expressions:  
\begin{eqnarray}
\hat{Q}_{q}(\tilde{\psi}) & \equiv &
Q_{c}(\langle\hat{q}\rangle_{\psi},\langle\hat{p}\rangle_{\psi}) 
+
\frac{\partial Q_{c}}{\partial p}  
(\hat{p} -\langle\hat{p}\rangle_{\psi})    
+ 
\frac{\partial Q_{c}}{\partial q} 
(\hat{q} - \langle\hat{q}\rangle_{\psi}),\\
\hat{P}_{q}(\tilde{\psi}) & \equiv &
P_{c}(\langle\hat{q}\rangle_{\psi},\langle\hat{p}\rangle_{\psi}) 
+
\frac{\partial P_{c}}{\partial p}  
(\hat{p} -\langle\hat{p}\rangle_{\psi})    
+ 
\frac{\partial P_{c}}{\partial q} 
(\hat{q} - \langle\hat{q}\rangle_{\psi}).
\end{eqnarray}
Considering these nonlinear operators, it is instructive to compute their
commutator in the ordinary sense of linear quantum theory:
\begin{eqnarray*}
[\hat{Q}_{q}(\tilde{\psi}),\hat{P}_{q}(\tilde{\psi})]
& = &\frac{\partial Q_{c}}{\partial q}
\frac{\partial P_{c}}{\partial p}[\hat{q},\hat{p}]
+\frac{\partial Q_{c}}{\partial p}
\frac{\partial P_{c}}{\partial q}[\hat{p},\hat{q}]\\
& = & \left(\frac{\partial Q_{c}}{\partial q}
      \frac{\partial P_{c}}{\partial p} -
      \frac{\partial Q_{c}}{\partial p}
      \frac{\partial P_{c}}{\partial q}\right)[\hat{q},\hat{p}]\\
& = & i\hbar\{Q_{c},P_{c}\}(\tilde{\psi})\hat{I}
\end{eqnarray*}
Preservation of the canonical commutation relations at each point in quantal
phase space is thus equivalent to the familiar condition (\ref{classcanon})
that the functions $Q_{c}$ and $P_{c}$ should generate a finite classical
canonical transformation. 

Arguing in purely physical terms, one now sees that a decision to work from
linear operators $\hat{q}$ and $\hat{p}$ involves no loss of generality. The
case of nonlinear operators merely signifies our freedom to carry out any
desired classical canonical transformation prior to quantization via
the rule ({\ref{nlham2}). On a much deeper level this freedom may be traced
to the special property of quantal phase space that all equivalence classes
$\tilde{\psi}(q,p)$ can be generated out of any single representative class.
The nonlinear operators defined above are therefore merely ordinary linear
operators acting upon an automorphism of the  original label space
$(q,p)\in\mathbb{R}^{2}$. One might just as well choose to carry out that
automorphism upon the abstract label space of states as a prelude to carrying
the Weyl group action to this space. At the deepest level function space has
no preferred coordinatization.

\section{Poisson Brackets and Canonical Transformations}
The previous argument is suggestive of a general role for Poisson brackets.
To establish that this is the case let us now work directly from the rules 
(\ref{eq:quan}) and (\ref{nlham2}). To any pair of classical phase space
functions $f_{c}(q,p)$ and $h_{c}(q,p)$ we assign the operators:
\begin{eqnarray}
\label{fdef}
\hat{f}_{q}(\tilde{\psi}) & \equiv &
f_{c}(\langle\hat{q}\rangle_{\psi},\langle\hat{p}\rangle_{\psi}) 
+
\frac{\partial f_{c}}{\partial p}  
(\hat{p} -\langle\hat{p}\rangle_{\psi})    
+ 
\frac{\partial f_{c}}{\partial q} 
(\hat{q} - \langle\hat{q}\rangle_{\psi})\\
\label{gdef}
\hat{h}_{q}(\tilde{\psi}) & \equiv &
h_{c}(\langle\hat{q}\rangle_{\psi},\langle\hat{p}\rangle_{\psi}) 
+
\frac{\partial h_{c}}{\partial p}  
(\hat{p} -\langle\hat{p}\rangle_{\psi})    
+ 
\frac{\partial h_{c}}{\partial q} 
(\hat{q} - \langle\hat{q}\rangle_{\psi})
\end{eqnarray}
whose  observable values, in the sense of the rule (\ref{expect}), clearly
agree with the values of the original classical functions for all $\psi$. 

We wish now to examine the general role of the usual quantal commutator. To
do that one must work from the postulate (\ref{eq:quan}). Before doing so
it is helpful to notice that although
$$[\hat{f}_{q}(\tilde{\psi}),\hat{h}_{q}(\tilde{\psi})]
= i\hbar \{f_{c},h_{c}\}(\tilde{\psi})\cdot\hat{I},$$
it happens that
$$i\hbar \{f_{c},h_{c}\}(\tilde{\psi})\cdot\hat{I}
\ne i\hbar \{f_{c},h_{c}\}_{q}^{\hat{}}(\tilde{\psi}).$$
Thus the quantization, according to (\ref{nlham2}), of the classical Poisson
bracket is not equal to the commutator of the quantizations of the original
classical functions. This is a very important observation in respect of the
Groenwold--van Hove theorem\cite{groen} which asserts the non--existence of a
quantization prescription having the property that it should preserve the
Poisson bracket.

However, it is true in general that:
$$\langle[\hat{f}_{q}(\tilde{\psi}),\hat{h}_{q}(\tilde{\psi})]\rangle
= i\hbar\langle 
\{f_{c},h_{c}\}_{q}^{\hat{}}(\tilde{\psi})\rangle.$$
A consequence of this peculiarity is that there does not appear to be any
simple analogue of the Heisenberg equation of motion for {\em operators\/}.
However, there is a Heisenberg equation of motion for expectation values.

To obtain this, suppose that
\begin{equation}
\label{tdep}
i\hbar\frac{d}{dt}|\psi\rangle =\hat{h}_{q}(\tilde{\psi})|\psi\rangle 
\end{equation}
determines $\psi(t)$ and consider the expression
$$\frac{d}{dt}\langle \psi|\hat{f}_{q}(\tilde{\psi})|\psi\rangle
= \langle \dot{\psi}|\hat{f}_{q}(\tilde{\psi})|\psi\rangle
+ \langle \psi|\hat{f}_{q}(\tilde{\psi})|\dot{\psi}\rangle
+  \langle \psi|\frac{d\hat{f}_{q}(\tilde{\psi})}{dt}|\psi\rangle.$$
Application of (\ref{tdep}) now reduces this to
$$\frac{d}{dt}\langle \psi|\hat{f}_{q}(\tilde{\psi})|\psi\rangle
= \frac{1}{i\hbar}\langle
[\hat{f}_{q}(\tilde{\psi}),\hat{h}_{q}(\tilde{\psi})]\rangle
+
\langle \psi|\frac{d\hat{f}_{q}(\tilde{\psi})}{dt}|\psi\rangle,$$
as a general result valid for any $\tilde{\psi}$--dependence. The problem
lies now within the final term. To evaluate this we need to know how
$\tilde{\psi}$ enters. Using the definition (\ref{gdef}) one can readily
compute the result 
\begin{eqnarray*}
\lefteqn{\frac{d\hat{f}_{q}(\tilde{\psi})}{dt} = 
\left(\frac{\partial f_{c}}{\partial p}
\frac{d\langle\hat{p}\rangle_{\psi}}{dt}
+
\frac{\partial f_{c}}{\partial q}
\frac{d\langle\hat{q}\rangle_{\psi}}{dt}
\right)
-
\left(\frac{\partial f_{c}}{\partial p}
\frac{d\langle\hat{p}\rangle_{\psi}}{dt}
+
\frac{\partial f_{c}}{\partial q}
\frac{d\langle\hat{q}\rangle_{\psi}}{dt}
\right)}\hspace{4.2cm}\\
&&
+
\left(\frac{\partial^{2} f_{c}}{\partial^{2} p}
(\hat{p} - \langle\hat{p}\rangle_{\psi})
+
\frac{\partial^{2} f_{c}}{\partial p\partial q}
(\hat{q} - \langle\hat{q}\rangle_{\psi})
\right)\frac{d\langle\hat{p}\rangle_{\psi}}{dt}\\
&&
+
\left(\frac{\partial^{2} f_{c}}{\partial q \partial p}
(\hat{p} - \langle\hat{p}\rangle_{\psi})
+
\frac{\partial^{2} f_{c}}{\partial^{2} q}
(\hat{q} - \langle\hat{q}\rangle_{\psi})
\right)\frac{d\langle\hat{q}\rangle_{\psi}}{dt}.
\end{eqnarray*}
Computing the expectation value of this result, one finds that it is 
identically equal to zero, whatever the evolution of $\psi$.

Thus we obtain the rather striking result
\begin{equation}
\label{poisson}
\frac{d}{dt}\langle \hat{f}_{q}(\tilde{\psi})\rangle
=
\langle \frac{1}{i\hbar}
[\hat{f}_{q}(\tilde{\psi}),\hat{h}_{q}(\tilde{\psi})]\rangle 
= \langle \{f_{c},h_{c}\}^{\hat{}}(\tilde{\psi}) \rangle, 
\end{equation}
where it should be noted that expectation values are essential. The result
(\ref{poisson}) fails as an equality for operators.

In this manner one sees that the nonlinear quantization condition derived at
(\ref{nlham2}) actually recovers Dirac's formal rule of
replacement\cite{dirac} in the modified form  
$$\langle \{\bullet,\bullet\}\rangle \rightarrow \langle
[\bullet,\bullet]/i\hbar \rangle.$$ 
The inclusion of the angle brackets now makes this an exact result which
depends in no way upon the magnitude of $\hbar$, but which applies only in
the case of a quantization procedure of the form (\ref{nlham2}).

It follows from (\ref{poisson}) that the dynamical system defined by
(\ref{eq:quan}) and (\ref{nlham2}) enjoys a complete inclusion of the full
group of classical symmetries. Any $\hat{h}_{q}(\tilde{\psi})$ can be now
considered as the generator of a one parameter group of symplectic
diffeomorphisms on quantal phase space\cite{arnold}, the canonical
transformations of Hamiltonian classical mechanics.  However, there now
occurs, at least within this special dynamical system, the additional freedom
to include the full group of nonlinear canonical transformations rather than
just the linear ones as are currently employed in the traditional phase space
formulation of quantum mechanics\cite{wigsen}.
 
\section{Measurement Theory}
As one might anticipate, there are some difficult conceptual problems
associated with the introduction of ordinary measurement theoretic concepts
to this unusual version of classical mechanics. 

\subsection{Dispersion and the Uncertainty Principle}
For instance, Heisenberg's uncertainty principle
\begin{equation}
\label{up} 
\langle(\Delta\hat{q})^{2}\rangle\langle(\Delta\hat{p})^{2}\rangle 
\ge \hbar/4,
\end{equation}
where $ 
\langle(\Delta\hat{q})^{2}\rangle \equiv 
\langle (\hat{q} -\langle\hat{q}\rangle)^{2}\rangle$
and $
\langle(\Delta\hat{p})^{2}\rangle \equiv 
\langle (\hat{p} -\langle\hat{p}\rangle)^{2}\rangle$,  
must remain a valid mathematical statement about wavefunctions and operators.

However, the role of (\ref{up}) is very unclear within this framework. The
difficulty is that the quantization rule (\ref{nlham2}) does not permit one
to go beyond operators that are first--order in $\hat{p}$ and $\hat{q}$. It
seems that such problems are mainly conceptual, but it is deemed important to
highlight them. It seems that the problems posed by the existence of an
embedding of classical mechanics within a generalized quantum theory must
provide an excellent focus for deeper investigation of the orthodox
epistemological framework.

Given the above observation, one might now pursue the idea of taking the
following definition of a generalized operator dispersion:
\begin{equation}
\label{dispersion} 
\langle(\Delta\hat{q})^{2}(\tilde{\psi})\rangle 
\equiv \langle \hat{q}^{2}(\tilde{\psi}) \rangle - 
       \langle \hat{q}(\tilde{\psi}) \rangle^{2}
\;\;\mbox{and}\;\;
\langle(\Delta\hat{p})^{2}(\tilde{\psi})\rangle 
\equiv \langle \hat{p}^{2}(\tilde{\psi}) \rangle - 
       \langle \hat{p}(\tilde{\psi}) \rangle^{2}.
\end{equation}
It is not clear that this is the only definition possible, but it seems
natural enough. In making the above choice we are merely exploring the
question of what can go wrong if one tries to carry over other features of
quantum measurement theory, rather than just the rule (\ref{expect}).

Having made the above choice, there now occurs a rather suprising phenomenon.
Noting that $q^{2}$ and $p^{2}$ are quantized under the rule (\ref{nlham2})
as the nonlinear operators 
$$\hat{q}^{2}(\psi)  = \langle q\rangle_{\psi} ^{2} 
+ 2 \langle q \rangle_{\psi}(\hat{q} - \langle q\rangle_{\psi})
\;\;\mbox{and}\;\;
\hat{p}^{2}(\psi)  =  \langle p\rangle_{\psi} ^{2} 
+ 2 \langle p \rangle_{\psi}(\hat{p} - \langle p\rangle_{\psi}),$$
one discovers immediately that 
$$\langle\psi|\hat{q}^{2}(\psi)|\psi\rangle = 
(\langle\psi| q |\psi \rangle)^{2}
\;\;\mbox{and}\;\;
\langle\psi|\hat{p}^{2}(\psi)|\psi\rangle = 
(\langle\psi| p |\psi \rangle)^{2},$$
from which it follows that
$$\langle(\Delta\hat{q})^{2}(\tilde{\psi})\rangle = 0
\;\;\mbox{and}\;\;
\langle(\Delta\hat{p})^{2}(\tilde{\psi})\rangle =  0.$$
So the choice of the rule (\ref{dispersion}) leads to the curious situation
that we have a theory with finite $\hbar$ in a regime of quantization that
appears to have no operator dispersion!

\subsection{Nonlinearity and the projection postulate}
This fact clearly signals that there will be severe problems in attempting to
carry over standard measurement theory to our nonlinear theory. The source
of the problem appears to lie with the projection postulate of ordinary linear
quantum theory. This is implicit in the interpretation of dispersion.

According to this rule, the probability of a state $\psi$ making a stochastic
transition to the state $\phi$, the eigenstate of an instantaneous
Hamiltonian, is given by:
\begin{equation}
\label{project}
p(\phi|\psi) = |\langle \phi|\psi\rangle|^{2}.
\end{equation}
Invocation of this rule requires that $|\langle \phi|\psi\rangle|^{2}$ be an
invariant. Currently it is generally argued\cite{jordan} that this alone
rules out the possibility of nonlinear evolution. Central to that argument is
Wigner's theorem, which shows that the only continuous probability preserving
automorphisms of the space of states are the linear unitary transformations.
Hence it is commonly asserted that quantal evolution must {\em always\/} be
linear.

The primary difficulty with this conceptual standpoint is that the stochastic
process $\psi\mapsto \phi$ according to the rule (\ref{project}) is manifestly
not a linear unitary process. If there is no transition or measurement, the
possibility of one, and the fact that (\ref{project}) gives the correct
probabilities, certainly demands, via Wigner's theorem, that the evolution
be linear in that situation\cite{jordan}. However, this argument does not
demand that quantal evolution be linear {\em at all times\/}. In order to
invoke Wigner's theorem, we must assume that the quantity (\ref{project})
has a physical meaning. In doing so, we accept that there is a phase of
quantal evolution that is not linear and which has the stochastic description
given by (\ref{project}). It is the possibility of this {\em special\/} mode
of evolution which enforces linearity whenever the quantum state $\psi$ is not
undergoing a process described by (\ref{project}) and where we wish to allow
that it might. This subtlelty of the argument is rarely appreciated. When it
is\cite{jordan}, it is usually flagged by stating that linearity is enforced
whenever the system in question is {\em isolated\/}. Often that proviso is
ignored.

Ordinarily, one now lives with the dichotomy by saying that quantum mechanics
is mighty peculiar and one must therefore accept that there be two processes;
one which is deterministic, and linear, the other which is stochastic and
manifestly nonlinear\cite{qmt}. However, if we desire a consistency of
dynamics then there appears to be a clean split and the two rules do not
rest well with one another\cite{penrose}. 

When we consider nonlinear quantum theory, the interesting thing is that
(\ref{project}) can no longer be valid in the sense that it is used now. This
situation arises because the Hilbert space inner product is no longer a
dynamical invariant. We cannot then reserve the possibility that a process
like (\ref{project}) may occur at any instant, if we were to do that then
Wigner's theorem demands linearity. However, here one may perceive an outlet.
The logical structure of quantum mechanics, as we are prepared to live with
it, does not appear to preclude a trichotomy of evolutionary processes. One
might just as well append nonlinear evolution as another special circumstance
(a facetious standpoint); or one might discover that it governs the emergence
of the special behaviour embodied in the rule (\ref{project}), see
Pearle\cite{pearle}. The first option one can resolve by
experiment\cite{jones3}, the second is interesting primarily from a
philosophical viewpoint.

\subsection{A speculation on the process of measurement}
The existence of our classical model certainly poses some very deep questions
about the degree to which the orthodox probabilistic interpretation of linear
quantum theory can be considered natural to the generalized dynamical
framework that it inhabits. Whereas rule (\ref{expect}) proves useful to both
linear quantum mechanics and our quantal model of classical mechanics, the
rule (\ref{project}) does not. One might argue that the success of our
program was a mere accident, but the mathematical result that is responsible
for it appears far from accidental\cite{weyl}.

Whereas the two theories of classical and quantal mechanics are generally
considered to be wholly incompatible; we are now led to the view that the
ultimate source of that incompatibility must lie within the projection
postulate. Our view is that the probabilistic interpretation of $\psi$ can
never be explained using the deterministic linear evolution dictated by the
ordinary Schr\"{o}dinger equation. If one desires an explanation, then we
note that the rule (\ref{project}) cannot survive as a possible instantaneous
stochastic process within any nonlinear phase of evolution. Considering these
facts, it seems that a resolution of the issue may creep by if one is
prepared to consider the curious possibility that the classical regime could
actually be a nonlinear regime of quantum theory.

Copenhagen orthodoxy characterises measurements as {\em special\/} processes 
involving metastable states of a macroscopic apparatus, which must always be
described in classical terms\cite{qmt,bohr}. We have seen that exact classical
mechanics requires a nonlinear quantum theory. Now suppose, for the sake of
argument, that classical mechanics actually is an effective nonlinear form of
quantum theory appropriate to macroscopic objects. Precisely how we are not
sure, but the clue is that no physical system describable in classical terms
could ever be considered {\em isolated\/} in the sense we understand that term
to apply within linear quantum theory as defining the circumstances under
which Wigner's theorem rules its decree. We already know that linear quantum
theory is accurate for the dynamics of microscopic systems. Our primary
supposition now demands some blend in between\cite{jones2}.

Let us now focus upon (\ref{project}) as representing the major logical
difficulty in our current understanding of quantum mechanics. The rule
itself works so we accept it. Consider now its invocation. We use it primarily
in those instances where the macroscopic world is coupled to the microscopic
world in a manner such that the latter may induce a registerable and
permanent change in the former\cite{bohr}. Given our outlandish premise that
classical mechanics may be an exact effective theory, it now appears
self--evident that any process which couples a macroscopic device to a
microscopic system would have to lie within the nonlinear regime of the
presumed generalized theory. Since the only manner in which we are able to
observe the microscopic regime is via such coupling, then it appears that we,
meaning physicists and their instruments of observation, knowing that we are
classical beings, must only ever have a nonlinear interaction with the
subatomic world. When sufficiently isolated, the microscopic world may quite
happily evolve in a linear fashion until such time as nonlinear creatures
such as ourselves try and observe it. Since experience tells us that
(\ref{project}) applies in just these circumstances, then one is led to look
for its motivation as an emergent property of quantal nonlinearity. 

The idea is striking. It seems to us to be the only way one might ever
resolve the difficulties offered by (\ref{project}). It is plausible because
(\ref{project}) has a generic mathematical form (it is always an inner
product\cite{jones3}). Were that not the case then there could be no universal
trick that might accomplish such a feat. Such a trick appears necessary.
However, granted the viewpoint, there is no obvious way to proceed. One would
have to search for such a result in the belief that it does exist, when it
may not. In this respect we are sad to admit that there appears to be very
little hope for it\cite{bell}. 

\section{Phase Dynamics and Anholonomy Effects}
Although we have lost operator dispersion as a signature of the quantal
origins of this version of classical mechanics, it is thoroughly remarkable
that in losing that quantal feature, classical mechanics actually gains for
itself a new and quite surprising one. Within this version of classical
mechanics there is now a natural $U(1)$ quantal phase factor associated to
any classical trajectory $(q(t),p(t))$. Classical mechanics acquires a
natural {\em geometric phase\/}. It is here that the role of $\hbar$ finally
shows up. We shall demonstrate that the embedded quantal geometric phase, in
combination with the dynamical phase, actually returns the {\em classical
action\/} divided by $\hbar$. 

The existence of a semiclassical interpretation of the Heisenberg--Weyl group
phase factor has been noted before\cite{perel}. In this connection, it is
already well--known, from semiclassical calculations, that Berry's adiabatic
quantal phase\cite{berry1} enjoys an intimate relationship with Hannay's
adiabatic classical angle\cite{hann,berry2}. Given that there is a good
understanding of that topic, our discussion shall focus upon the classical
analogue of Aharanov and Anandan's\cite{ahar} non--adiabatic version of
Berry's phase. 

In our previous correction of the energy, we upset the phase dynamics in
a most interesting way. A simple calculation, performed in appendix two, now
reveals that the leading phase associated with any state $\phi_{0}$ should be
readjusted to read  
\begin{equation}
\label{classtraj} 
|q(t),p(t);\phi_{0}\rangle = e^{i(\gamma(t) -\beta(t))}
\hat{U}[q(t),p(t)]|\phi_{0}\rangle,
\end{equation} 
where the explicit formul{\ae} for the phases are:
\begin{eqnarray} 
\label{AAphase} \gamma(t) & = & \frac{1}{\hbar}
\int_{0}^{t} \left(\frac{\dot{q}p -\dot{p}q}{2}\right) d\tau,\\
\label{dynamic}
\beta(t)  & = & \frac{1}{\hbar}
\int_{0}^{t} H_{c}(q,p) d\tau.
\end{eqnarray}
Observe that (\ref{classtraj}) contains the two usual contributions to the
overall phase. This decomposition amounts to an application of the now
standard technique for calculating the intrinsic phase anholonomy for a
wavefunction circuit\cite{phase}, as first discovered by Berry in his now
famous paper\cite{berry1}. The first phase $\gamma(t)$ has a purely geometric
origin, we recognise it as the Aharanov--Anandan form of Berry's phase. The
second phase $\beta(t)$ is the usual dynamical phase. 

As one might expect $\gamma(t)$ has, in this one degree of freedom instance, 
a very simple geometrical interpretation as the integrated sectorial velocity
(the rate of change of swept area in phase space, as measured from the
origin, see Fig. \ref{fig4}). Significantly the area change has a sign
attached which makes the swept area positive for rotation in a clockwise
sense about the origin. On closed phase space circuits $\Gamma$, of arbitrary
shape, we therefore expect a relation between the $\gamma(\Gamma)$
accumulated on a loop and the area enclosed by that loop. Taking account of
the implicit sign convention one finds that:    
$$ \int_{0}^{T} p\dot{q}\,dt =
+\oint_{\Gamma} p\,dq \;\;\mbox{and}
   \int_{0}^{T} q\dot{p}\,dt = -\oint_{\Gamma} p\,dq,$$
where $T$ is the time taken to execute one circuit. The negative sign
arises in the second case because the sense of traversal is then that 
of a negative area contribution. The total result is 
\begin{equation} 
\gamma(\Gamma) = +1/\hbar \oint_{\Gamma} p\,dq.  
\end{equation}
So that one recognises the old Bohr--Sommerfeld quantization rule\cite{bohr},
$$ \oint_{\Gamma} p\,dq = 2\pi n \hbar,$$
in a new light as a constraint that $\gamma(\Gamma)$ should be an integral
multiple of $2\pi$. 
%
%

The total phase $\phi= \gamma -\beta$ can now be recast in the most
interesting form: 
\begin{equation}  
\phi(\Gamma) = \frac{1}{\hbar}\oint_{\Gamma}
p\,dq -  \frac{1}{\hbar}\oint_{\Gamma} H(q,p)\,dt = 
\frac{1}{\hbar} \int_{0}^{T} L\,dt 
\end{equation}
which one immediately recognizes as the classical action. This relationship
suggests that quantal geometric phases upon closed loops might well be
interpreted as the natural action variables of quantum mechanics. 

Note that the correspondence of $\gamma(t)$ to the abbreviated classical
action, $S = \oint p\,dq$, is confined to closed trajectories. Upon open
trajectories the derivatives of both differ by a term of the form $1/2
d/dt(pq)$. One can trace this phenomenon to an essential arbitrariness
connected with the manner in which one might choose to close any given open
trajectory in phase space. We intend to return to deeper consideration of this
question in a subsequent publication. At this stage it is instructive to
conclude this section with two incidental comments.

Given our nonlinear quantization condition, one can now model any $n$--degree
of freedom classical system using the creation and annihilation operators for
$n$--degrees of freedom\cite{perel}. These have the standard commutation
relations:  
$$[\hat{a}_{j},\hat{a}^{\dagger}_{k}] = \delta_{jk} \;\;j,k\in[1,n].$$
Under this extension, the Weyl operators generalize to
$$D({\bf \alpha}) \equiv \exp \left
\{\sum_{j=1}^{n}\alpha_{j}\hat{a}^{\dagger}_{j} - 
\alpha_{j}^{*}\hat{a}_{j}\right\},$$
with the obvious multiplication rule
$$D({\bf \alpha})D({\bf \beta})= \exp \left
\{\sum_{j=1}^{n} {\rm Im}[\alpha_{j}\beta^{*}_{j}]\right\}
D({\bf \alpha + \beta}).$$
Repeating an essentially identical argument one recovers the $U(1)$ phase
$$\dot{\gamma}(t) = -{\rm Im}\,[\dot{\bf \alpha}\cdot{\bf \alpha}^{*}].$$ 
This expression appears significant in respect of a throwaway remark made
by Berry in his introductory article\cite{berry3}. 

In the language of that paper the above equation amounts to the assignment of
a semiclassical phase $\gamma(\Gamma)$ on closed curves $\Gamma$ bounding a
surface $S$, where this is given by the flux of a classical {\em 2--form},
$d{\bf p}\wedge\cdot d{\bf q}$, as the simple expression    
\begin{equation}   
\gamma(\Gamma) =  +
\frac{1}{\hbar}\int\int_{\partial S =\Gamma} d{\bf p}\wedge\cdot d{\bf q}. 
\end{equation} 
However, unlike the situation described in\cite{berry3}, this is not the angle
averaged flux for an adiabatic excursion. In this situation we are dealing
with the nonadiabatic phase. This means that the $d$'s now link ${\bf p}$ and
${\bf q}$, so that the above relationship involves actual phase space forms
rather than the parameter space forms of Berry's article\cite{berry3}. This
is interesting because it suggests the possibility that one might take a 
direct approach to the adiabatic phase as being an appropriately averaged
swept area change in phase space due to the cyclic deformation of the phase
space trajectories themselves.

Our second comment concerns the fact that the expression 
$$\gamma(\Gamma) = - {\rm Im} \oint_{\Gamma} {\bf \alpha}^{*}\,d{\bf \alpha}$$
is already a Poincar\'{e} integral invariant\cite{arnold}. In terms of
the classical variables $({\bf q},{\bf p})$ it is the circuit integral
$$\gamma(\Gamma) = \frac{1}{\hbar}\oint_{\Gamma} 
\sum_{j=1}^{n} p_{j}\,dq_{j},$$ 
which is equal to the area integral given previously. 

This observation demonstrates that the quantal version of an integrable
classical system, as we have defined it, will have $n$ geometric phase
``actions'' associated with the $n$ irreducible contours that exist upon any
``torus'' in the $n$--dimensional complex quantal phase space obtained via
the generalized projection map $\Pi[\psi(t)]= ({\bf q}(t),{\bf p}(t))$.
Considering the evolution of $\psi(t)$ itself, we could say that it explores
an $n$--dimensional sub--manifold of the infinite dimensional Hilbert space,
which manifold might then be called a functional torus.

\section{Solutions of the Classical Equation}
We shall now exhibit a particular example of the classical Schr\"{o}dinger
equation in a familiar explicit representation. It is important to
understand that this is just one example of what happens to be a perfectly
general result. 

For the one dimensional classical Hamiltonian,
$$H_{c}(q,p) = \frac{p^{2}}{2m} + V(q),$$
our corrected quantization prescription (\ref{nlham2}) gives the result
\begin{equation}
\hat{H}_{q}(\tilde{\psi})  = 
\frac{\langle \hat{p} \rangle^{2}}{2m} + V(\langle q
\rangle)  + \frac{\langle \hat{p} \rangle}{m}
(\hat{p} - \langle \hat{p} \rangle) +
\frac{\partial V}{\partial q}(\langle q\rangle) (\hat{q} -\langle \hat{q} \rangle).
\end{equation} This abstract operator Hamiltonian can now be carried into the
standard Schr\"{o}dinger representation via the usual rules:
$$\hat{q}\mapsto q,\;\;\hat{p}\mapsto
-i\hbar\frac{\partial}{\partial q},\;\;\mbox{and}\;\;|\psi\rangle\mapsto
\psi(q).$$  
In the coordinate representation the required equation now corresponds to a
nonlinear first-order integrodifferential equation of the explicit form: 
\begin{equation}
\label{class}
i\hbar \left(\frac{\partial }{\partial t}
+            \frac{{\langle \hat{p} \rangle}}{m} 
             \frac{\partial }{\partial q}\right) \psi(q,t) = 
\left (V(\langle \hat{q} \rangle) 
+ \frac{\partial V}{\partial q}(\langle \hat{q} \rangle)  
(q - \langle \hat{q} \rangle)  
-\frac{\langle \hat{p} \rangle^{2}}{2m} \right) \psi(q,t),
\end{equation}
where of course $\langle \hat{q} \rangle$ and  $\langle \hat{p} \rangle$ must at
all times satisfy the relations
$$ \langle \hat{q} \rangle = \int_{-\infty}^{\infty}\!
q\psi^{*}(q,t)\psi(q,t)\,dq
\;\; \mbox{and} \;\;
\langle \hat{p} \rangle = \int_{-\infty}^{\infty}\!
-i\hbar\psi^{*}(q,t)\frac{\partial}{\partial q}\psi(q,t)\,dq.$$
The equation (\ref{class}) is most unusual. At first sight the prospect of
attempting to solve such a highly nonlinear equation might appear forbidding.
However, here we know that the solutions of (\ref{class}) are characterised by
the fact that the position and momentum expectation values must follow
precisely the classical trajectories. 

The idea is then to seek solutions $\psi(q,t)$ that are parametrised by 
the purely numerical functions of time
\begin{equation}
\label{constraint} 
Q(t) \equiv \langle \psi(t) | \hat{q} | \psi(t) \rangle
\;\; \mbox{and} \;\;
   P(t) \equiv \langle \psi(t) | \hat{p} | \psi(t) \rangle.
\end{equation}
From our original Weyl operator considerations, we already know that the
complete evolution operator solution of (\ref{class}) must be
$$\tilde{U}[Q(t),P(t)] = e^{i\phi(t)} U[Q(t),P(t)],$$
where $Q(t)$ and $P(t)$ are parameters obtained via solving the purely
classical equations of motion 
\begin{equation} 
\label{cond}
\frac{dP}{dt}  = 
 - \frac{\partial V}{\partial q} (Q)\;\;\mbox{and}\;\;
\frac{dQ}{dt}  =  
+ \frac{P}{m}
\end{equation}
and the phase $\phi(t)$ is fixed via the group anholonomy effects
considered in the preceding section to be
\begin{equation}
\label{phase2}
\phi(t) = \frac{1}{\hbar}\int_{0}^{t}
\left[\frac{1}{2}\left(
P(\tau)\dot{Q}(\tau)  -
Q(\tau)\dot{P}(\tau) \right)
- \frac{P(\tau)^{2}}{2m} - V(Q(\tau))
\right]\,d\tau.
\end{equation}
Using this insight we can now construct a general family of solutions to
(\ref{class}) and verify the condition (\ref{cond}) as an essential
constraint upon the time--dependent parameters $Q(t)$ and $P(t)$.

From \S {\rm III} C, equation (\ref{antiord}), recall that $U[Q(t),P(t)]$ has
the convenient coordinate representation
$$U[Q(t),P(t)] = 
\exp\left\{-\frac{iP(t)Q(t)}{2\hbar}\right\} 
\exp\left\{+ \frac{iP(t) q}{\hbar}\right\}
\exp\left\{- Q(t)\frac{\partial}{\partial q}\right\}.$$
Including the phase (\ref{phase2}), we can now construct the required 
parametric family of solutions as the states
\begin{equation}
\label{answer}
\psi(q,t) \equiv e^{i\phi(t)} 
\exp\left\{-\frac{iP(t)Q(t)}{2\hbar}\right\} 
\exp\left\{+ \frac{iP(t) q}{\hbar}\right\}  
\psi_{0}(q - Q(t)),
\end{equation}
where the time dependence of these enters purely via the group parameters
$Q(t)$ and $P(t)$. Provided that $\phi_{0}(q)$ has both expectation values
equal to zero, it follows that the constraint (\ref{constraint}) is at all
times trivially satisifed. 

The idea is now to substitute  the trial solution (\ref{answer}) into the 
wave equation (\ref{class}) so as to discover a necessary condition upon
the parameters $Q(t)$ and $P(t)$. After some suppressed algebra, during 
which one must be careful to note that the special form of (\ref{answer})
enforces the equalities (\ref{constraint}), one arrives at the required
condition: 
$$i\hbar\left(\frac{P(t)}{m} -\frac{dQ(t)}{dt}
\right) \psi_{0}'(q -Q(t)) = 
\left(\frac{\partial V}{\partial q}(Q(t))
- \frac{dP(t)}{dt}\right) (q - Q(t))
\psi_{0}(q -Q(t)),$$
in which it should be noted that some phase factors have been conveniently
cancelled. Examination of the above equation now shows that equality for all
equivalence class representatives $\psi_{0}(q)$ demands that $Q(t)$ and
$P(t)$ should satisfy Hamilton's equations (\ref{cond}), as claimed. 

So (\ref{class}) actually admits an infinite family of physical travelling
wave solutions. Moreover, the complete family of solutions must exhaust the
set of all square integrable and appropriately differentiable representative
functions $\psi_{0}(q)$. It is trivially assured that not one of these
solutions exhibits dispersion phenomena of any kind, they simply follow the
classical trajectories via the classical evolution of the functional
parameters $Q(t) = \langle \hat{q}\rangle$ and $P(t) = \langle
\hat{p} \rangle$.

To consider a particular elementary example, one might set $V(q)=0$ so
as to obtain the free particle equation
$$i\hbar\left(\frac{\partial }{\partial t}
+            \frac{{\langle \hat{p} \rangle}}{m} 
             \frac{\partial }{\partial q}\right) \psi(q,t) = 0.$$
For the initial conditions $Q(0) = Q_{0}$ and $P(0) = P_{0}$ this equation
is solved by the family 
$$\psi(q,t) = \exp^{iP_{0}q/\hbar}\phi_{0}(q -  P_{0}t/m - Q_{0}),$$
with $\phi_{0}$ any state belonging to $\tilde{\psi}(0,0)$.

\section{Recovering Standard Quantum Mechanics}
We come now to the most remarkable result of all. Recall that the form of
the corrected energy Hamiltonian (\ref{nlham2}):
\begin{equation}
\label{lin}
\hat{H}_{q}(\tilde{\psi})\equiv H_{c}
+ \frac{\partial H_{c}}{\partial p} 
(\hat{p} -\langle\hat{p}\rangle_{\psi})   
+ \frac{\partial H_{c}}{\partial q} 
(\hat{q} - \langle\hat{q}\rangle_{\psi}),
\end{equation}
was constrained by the twin demands of a dynamical and energetic classical
correspondence such that the position and momentum expectation values should
track precisely the trajectories of any chosen classical Hamiltonian system.

In looking at the form of (\ref{lin}) one is naturally led to wonder about
that theory which corresponds to the continuation of the implicit Taylor
series to all orders. To investigate this possibility we require a way to
write such an infinite operator valued series in compact form. As a simple
way to explore that idea we elect to write the result in terms of a 
$\psi$--dependent {\em quantization operator\/}, which we define as
follows\cite{jones2}:   
\begin{equation}
\label{quant}
\hat{\cal Q}_{\psi} \equiv 
\exp \left\{\sum_{k=1}^{n} 
(\hat{q}_{k} - \langle
\hat{q}_{k}\rangle_{\psi})
\frac{\partial}{\partial q_{k}} + 
(\hat{p}_{k} - \langle \hat{p}_{k}\rangle_{\psi})
\frac{\partial}{\partial p_{k}}\right\}. 
\end{equation}
Under consideration is a classical system with $n$ degrees of freedom and an
associated quantal system having $n$ canonically conjugate pairs of position
and momentum operators $\hat{q}_{j}$ and $\hat{p}_{j}$. The idea is that we
should achieve the process of quantization via application of (\ref{quant})
to the classical function $f_{c}({\bf q},{\bf p})$ so as to arrive at the
quantal operator $f_{q}({\bf \hat{q}},{\bf \hat{p}})$. 

Recognising that the ordinary $c$-number operational calculus permits us
to write the identity
$$\exp\{(z - z_{0})\frac{\partial}{\partial z}  +
(w - w_{0})\frac{\partial}{\partial w}\}f(z_{0},w_{0})
= f(z,w),$$
when it is understood that all partial derivatives are evaluated at the point
$(z_{0},w_{0})$, we are led to define the action of (\ref{quant}) so that all
$c$-number derivatives are evaluated in like manner at the quantal expectation
values. Proceeding in this intuitive fashion one writes the expression
\begin{equation} 
\hat{\cal Q}_{\psi}\circ
f_{c}(\langle\hat{q}_{1}\rangle_{\psi},\ldots,
\langle\hat{q}_{n}\rangle_{\psi};
\langle\hat{p}_{1}\rangle_{\psi},\ldots,
\langle\hat{p}_{n}\rangle_{\psi})
= 
\hat{f}(\hat{q}_{1},\ldots,\hat{q}_{n}; \hat{p}_{1},\ldots,\hat{p}_{n}).
\end{equation}
where it is expected that the formal properties of the implicit Taylor series
shall ensure that the operator replacement is achieved in a manner that is
independent of the choice made for the $\psi$ expansion. We shall not
prove this here, reserving that for a fuller treatment elsewhere.   

For the present purpose, it is clear that (\ref{quant}) is unique, and that it
is a {\em linear\/} operator--valued map on a functional domain. Linearity
of this mapping is here meant in the sense that
$$\hat{\cal Q}_{\psi}\circ[f + g] = \hat{\cal Q}_{\psi}\circ f + 
\hat{\cal Q}_{\psi}\circ g,$$
although it happens that the result is also a linear operator.

Given that we are to understand (\ref{quant}) in the ordinary sense of a
differential operator power series expansion
$$\exp {\cal D} \equiv 1 + {\cal D} + 1/2! {\cal D}^{2}\ldots,$$ 
where the unusual feature of ${\cal D}$ is simply the fact that it now
carries within it some $q$--number coefficients, it is clear that the chosen
mapping must achieve a replacement of the arguments of the classical
Hamiltonian by canonically conjugate operators in a manner that automatically
implements Weyl's symmetric ordering of the troublesome noncommuting factors,
as generated by the cross derivatives of $q_{k}$ and $p_{k}$. To cement the
connection with canonical quantization, we remark that the ${\cal Q}_{\psi}$
prescription can be shown to be equivalent to Weyl's operator Fourier
transform\cite{weyl}. As one might expect, given that equivalence, it is as
well to mention also that one can use ${\cal Q}_{\psi}$ to rederive Moyal's
bracket\cite{moyal} as the correct phase space analogue of the quantal
commutator. Proofs shall be given in a forthcoming publication.

For the purpose of this article, the assertion is that ${\cal Q}_{\psi}$
transforms any classical phase space function into a quantal operator
equivalent, and that it does so in a manner which agrees with the standard
canonical quantization procedure on cartesian coordinates, as augmented by
the Weyl ordering rule. In this respect, (\ref{quant}) has the important
advantage that the procedure for operator replacement is now transparent. 

\subsection{An example: the generalized harmonic oscillator}
Since we here omit a full treatment of (\ref{quant}), a concrete example is 
useful to illustrate these points. Consider the ubiquitous one--dimensional
generalized Harmonic oscillator: 
$$H_{c}(q,p) = \frac{1}{2}(a p^{2} + b pq + c q^{2}).$$ 
There are two possible quantized versions according to the chosen order  of
approximation. The nonlinear first--order version is 
\begin{eqnarray*}
\hat{H}_{q}(\tilde{\psi}) 
& = & 1/2(a \langle \hat{p} \rangle^{2}
 + b \langle \hat{p} \rangle \langle \hat{q} \rangle 
+ c \langle \hat{q} \rangle ^{2}) \\
& & +  a \langle \hat{p} \rangle (\hat{p} - \langle \hat{p} \rangle) \\
& & + 
b/2
(\langle \hat{p} \rangle (\hat{q} - \langle \hat{q} \rangle)
+\langle \hat{q} \rangle (\hat{p} - \langle \hat{p} \rangle)\\
& & +  c \langle \hat{q} \rangle (\hat{q} - \langle \hat{q} \rangle)
\end{eqnarray*} 
This truncation involves neglect of the second--order contribution
$$ a/2 (\hat{p} - \langle \hat{p} \rangle)^{2}
+  b/2[(\hat{p} - \langle \hat{p} \rangle)( \hat{q} - \langle \hat{q} \rangle)
  +  (\hat{p} - \langle \hat{p} \rangle)( \hat{q} - \langle \hat{q} \rangle)]
+  c/2(\hat{q} - \langle \hat{q} \rangle)^{2}.$$
Inclusion of the above term now recovers, in a natural fashion,
the standard $\psi$--independent and correctly symmetrised result    
$$\hat{H}_{q} = \frac{1}{2}(a \hat{p}^{2}  + b/2[\hat{p}\hat{q} +
\hat{q}\hat{p}]  + c \hat{q}^{2}).$$
In comparing the possible truncations of the operator valued Taylor series
it is instructive to compute the energy expectation values. Doing that
one discovers that the first--order version yields the correct classical
energy:
$$E_{c} = \frac{1}{2}(a \langle \hat{p}\rangle^{2} 
+ b\langle \hat{p} \rangle \langle \hat{q} \rangle 
+ c \langle \hat{q} \rangle^{2}).$$ 
Upon inclusion of the second--order term, there appears a zero
point contribution and one obtains the total result:
$$E_{q} = \frac{1}{2}(a \langle \hat{p}^{2}\rangle 
+ b/2\langle \hat{p}\hat{q} + \hat{q}\hat{p} \rangle  
+ c \langle \hat{q}^{2} \rangle).$$
As this example demonstrates, the classical Schr\"{o}dinger equation can be
viewed as representing a natural approximation to the full quantum theory. Any
power series will do to represent the full quantal Hamiltonian. However, when
considering the truncated series it is natural to adapt the approximation to
the particular $\psi$ in view. The successive truncations of a power series
expanded about the expectation values of the instantaneous state could well
be expected to provide a fair approximation to the correct quantum 
dynamics\cite{mess}. The two noteworthy features of this approach
are that the first order approximation actually returns exact classical
mechanics and that this approximation corresponds to a nonlinear dynamics. 

\section{Conclusion}
We began this paper by posing a simple question that induced us to construct 
a rather odd looking quantal dynamics based solely upon the desire to see
classical behaviour mirrored in the evolution of entire classes of quantal
wavefunctions. Remarkably, the resulting dynamical structure was fixed by
this approach. Furthermore, the success of the procedure exposes the crucial
significance of the linear action of the Heisenberg--Weyl operator. The
remarkable mathematical property of this group, namely that it provides a ray
representation of the Abelian group of translations upon the plane, may be
considered to be directly responsible for the success of our entire program.
It also explains why classical mechanics acquires a phase factor. It seems
that one need only adjoin the postulate that the fundamental kinematical group
of motions upon the classical phase plane should have a projective structure
and the entire result follows through of necessity\cite{weyl}.

Concerning the perceived connection with Weinberg's theory\cite{wein1,wein2},
it is interesting that both structures employ norm--preserving dynamics. In
the case of Weinberg's work there is an insistence upon the more restrictive
property that the dynamics, and all observables, should remain unchanged under
scaling of $\psi$ by any non zero complex $Z$. This restriction to rays only
can be incorporated into our structure by exploiting the norm--preserving
property of our theory, so as to replace all appearances of $\psi$ in the
arguments of our nonlinear operators by $\psi/n^{1/2}$, where $n$ is the
preserved norm $n = ||\psi||^{2}$. An observation of this kind led us to
discover a natural Weinberg analogue of our quantal model of classical
mechanics\cite{jones}. We shall now describe the link between the present
paper and that paper.

Considering the fact that our Hamiltonian (\ref{nlham2}) yields an energy
observable with the classical value
$H_{c}(\langle\hat{q}\rangle,\langle\hat{p}\rangle)$ 
and noting that the central objects of Weinberg's theory are observables
$h(\psi,\psi^{*})$ that are homogeneous of degree one in both $\psi$ and
$\psi^{*}$ we were led to attempt to carry our result directly into his
theory. The obvious way to do this is to first generalize Weinberg's
$\star$--algebra of observables\cite{wein1} to its natural
infinite--dimensional analogue, and to then make the ansatz:  
$$h(\psi,\psi^{*}) = n\cdot 
H_{c}(\langle\hat{q}\rangle,\langle\hat{p}\rangle),$$  where the homogeneity
requirement is catered for by the revision $$\langle\hat{q}\rangle_{\psi} 
= \langle \psi|\hat{q}|\psi\rangle/n
\;\;\mbox{and}\;\;
\langle\hat{p}\rangle_{\psi} 
= \langle \psi|\hat{p}|\psi\rangle/n.$$
In the paper\cite{jones} we pursued this route as an entry point to Weinberg's
theory and uncovered a precisely analogous result. This situation provides
very strong evidence for the conjecture that the two dynamical structures are
ultimately equivalent. Given that the logical development of both approaches
rests upon independent footing, one is inclined to view the territory under
survey as being both rich and deep\cite{deep}.

In conclusion, it has to be said that the author was surprised by the success of this simple program. Elsewhere, we explored an interpolative scheme of quantal nonlinearity\cite{jones2} to explore the hypothesis of a {\em mesoscopic quantum dynamics\/}. However, the idea led to a free theory which is not falsifiable. Such theories are sterile for any new physics.

While the resulting system of interpolative dynamics is stimulating for mathematical studies, it calls attention to a central
problem with the oft-mentioned notion of a {\em classical domain\/} as distinct from a {\em quantum domain\/}. When 
this notion is taken seriously, it soon leads to a raft of logical contradictions. One can show that a controllable non-entangling 
mode of physical interaction would  violate the uncertainty principle\cite{exclusion}.

In virtue of this no-go theorem, the Classical Schr\"{o}dinger Equation constructed here is properly viewed as an example of an effective theory or {\em approximate dynamical system\/}. One can view it as the natural result of freezing out the quantum degrees of freedom through neglect of higher--order terms in the operator Taylor-series expansion.

In connection with any fundamental non--linearity, the no-go theorem suggests that we look for a class of non--linear terms which are not controllable, by hypothesis. One possibility is a non--linearity due to physical self--interaction\cite{lor}. One definite
proposal has since been elaborated with what we believe to be a self--consistent physical  interpretation\cite{nqg}.

\section{Acknowlegments}
The majority of this work was completed at the University of Melbourne, some twenty years ago. However, the
author chose not to publish without a clear physical interpretation. That was provided some years later, but by
then the author had temporarily lost the text. Subsequently, the text was found again and restored from  
electronic media. It appears here, modified in light of the later physical interpretation.

For the original work, I am grateful to both Professors B.H.J. McKellar and A.G. Klein for providing financial support while the research was carried out. I thank Prof. I.C. Percival,  Dr. N.E. Frankel, Dr. A.J. Davies, Dr. S. Kuyucak, Dr. V. Kowalenko and Dr. R. Volkas for helpful discussions. I also thank Prof. S Weinberg for his generous hospitality at the University of Texas at Austin, and for helpful discussions about our respective mathematical results and the delicacy of their physical interpretation.

\clearpage
\appendix{Derivation of the Hamiltonian}
There is a very nice property of the Schr\"{o}dinger equation, when
cast in the evolution operator form
$$i\hbar \frac{d}{dt} \hat{U}(t,t_{0}) = \hat{H}(t) \hat{U}(t,t_{0}).$$
This concerns the fact that evolution operators are the natural
analogues of phase space trajectories in the following special sense.
Given any curve on the group manifold, namely an evolution operator
trajectory $\hat{U}(t,t_{0})$, we can deduce the quantal Hamiltonian
appropriate to that trajectory. 

Consider the group trajectory $\hat{U}(t,t_{0})$. Applying only the
essential unitarity constraint one can readily deduce the result
\begin{equation} 
\hat{H}(t) \equiv  i\hbar [\frac{d}{dt} \hat{U}(t,t_{0})]
\hat{U}(t,t_{0})^{\dagger}
\end{equation}
as the defining equation for the Hamiltonian $\hat{H}(t)$.

Of interest here are group trajectories of the form
$$\tilde{D}(\alpha(t))\equiv e^{i\phi(t)} D(\alpha(t))$$
where $D(\alpha(t))$ is a member of the Heisenberg--Weyl group. The
multiplication rule for this group is entirely sufficient to determine
the necessary operator derivative, where this is defined as follows
\begin{equation}
[\frac{d}{dt} \tilde{D}(\alpha(t))]
\tilde{D}(\alpha(t))^{\dagger}  =  \lim_{\delta t \rightarrow 0}
\frac{\tilde{D}(\alpha(t) + \dot{\alpha}(t)\delta t)
\tilde{D}(\alpha(t))^{\dagger} - \hat{I}}{\delta t}.
\end{equation}
Using the fact that 
$$\tilde{D}(\alpha(t))^{\dagger} = e^{-i\phi(t)} D(-\alpha(t))$$
one rewrites the above expression as
\begin{equation}
[\frac{d}{dt} \tilde{D}(\alpha(t))]
\tilde{D}(\alpha(t))^{\dagger}  =  \lim_{\delta t \rightarrow 0}
\frac{e^{{+i(\dot{\phi} - {\rm Im}[\dot{\alpha}\alpha^{*}])\delta t}}
D(\dot{\alpha}(t)\delta t) - \hat{I}}{\delta t}. 
\end{equation}
Expanding to first--order in $\delta t$ then yields
\begin{equation}
[\frac{d}{dt} \tilde{D}(\alpha(t))]
\tilde{D}(\alpha(t))^{\dagger}
= +i(\dot{\phi} - {\rm Im}[\dot{\alpha}\alpha^{*}])\hat{I}
  +\dot{\alpha}\hat{a}^{\dagger} - \dot{\alpha}^{*}\hat{a}
\end{equation}
as the final result. Note that the arbitrariness of $\phi(t)$ shows up
here as a contribution $\dot{\phi}\hat{I}$ in the resulting Hamiltonian.
One choice is especially convenient, that which cancels the term
$-{\rm Im}[\dot{\alpha}\alpha^{*}]$. To this end set
$$\phi(t) = +\int_{0}^{\tau}{\rm Im}[\dot{\alpha}\alpha^{*}]\,dt$$ 
from which we obtain the Hamiltonian
$$-i\hat{H}(\alpha(t)) = 
\dot{\alpha}\hat{a}^{\dagger} - \dot{\alpha}^{*}\hat{a}.$$ 
This choice of phase now yields the propagator equation of motion
\begin{equation} 
\frac{d}{dt} \tilde{D}(\alpha(t)) = 
(\dot{\alpha}\hat{a}^{\dagger} -
\dot{\alpha}^{*}\hat{a})
\tilde{D}(\alpha(t)),
\end{equation}
which verifies the result (\ref{derivative}).

\appendix{Derivation of the Geometric Phase}
To derive the geometric phase appropriate to any classical trajectory 
we shall work first with the $D(\alpha)$ operators and then revert to
classical phase space coordinates at the end of the argument. 

Following the Aharanov--Anandan exposition\cite{ahar} of the geometric
phase as being associated with a path in the projective Hilbert space,
and noting that our propagators are always Weyl operators, one can consider 
an arbitrary initial state $\psi(0)$, and a propagator $D(\alpha(t))$,
as determining the ray space trajectory:
$$|\psi(t)\rangle = D(\alpha(t))|\psi(0)\rangle.$$
We can then calculate the geometric phase using the rule\cite{ahar}
\begin{equation}  
\dot{\gamma}(t) = i\langle \psi(0)| D(\alpha(t))^{\dagger}
[\frac{d}{dt} D(\alpha(t))] |\psi(0)\rangle. 
\end{equation}
This way of generating an arbitrary trajectory builds up increments in
the phase $\gamma(t)$ via the geodesic transport of $|\psi(0)\rangle$, out
along a straight line from the origin, along a path segment dictated by that
upon the group, and thence back to the origin\cite{sam}. 

Using essentially the same argument as in appendix one, the time
derivative of the phase is now easily calculated to be 
$$ \dot{\gamma} = - {\rm Im}[\dot{\alpha}\alpha^{*}],$$ 
whence we obtain the desired result 
\begin{equation}
\gamma(t) = 
- \int_{0}^{t}{\rm Im}[\dot{\alpha}\alpha^{*}]\,d\tau.
\end{equation}
It is instructive to use the substitution $\alpha = c_{q}q + ic_{p}p$
and $c_{q}c_{p} = 1/2\hbar$ so as to rewrite this phase as
\begin{equation} 
\gamma(t) = \int_{0}^{t}\frac{p\dot{q} - \dot{p}q}{2\hbar}\,d\tau.
\end{equation}
This assigns a natural semiclassical Aharanov--Anandan quantal geometric
phase to an arbitrary classical trajectory. Note that this phase is a
geometric property of the trajectory alone and the chosen criterion for
closing open trajectories. In this respect, other closing criteria are
possible, such as closing to the $q$ or $p$ axes, rather than to the origin.
These shall of course yield different values for the integrated phase upon
open trajectories. Moreover, the manner of closing is not preserved under
canonical transformations. However, upon naturally closed trajectories the
arbitrariness is removed and the geometric phase, being a well defined phase
space area, must be invariant under all classical canonical transformations.
The only phase freedom is then that of adding a multiple of the unit operator
to the Hamiltonian. This can be interpreted as an expression of the standard
freedom to change the energy origin in both classical and quantum mechanics.

To compute the overall phase, we must now subtract the dynamical contribution
$$\beta(t) = \frac{1}{\hbar}\int_{0}^{t}
\langle\psi|\hat{H}_{q}(\tilde{\psi})|\psi\rangle\, d\tau,$$
which, given the rule (\ref{nlham2}), is easily seen to be
$$\beta(t) = \frac{1}{\hbar}\int_{0}^{t}
H_{c}(q,p)\, d\tau.$$
In this manner one recovers equations (\ref{classtraj}), (\ref{AAphase})
and (\ref{dynamic}), as required.

\clearpage

\begin{figure}[p]
\includegraphics[width=80mm]{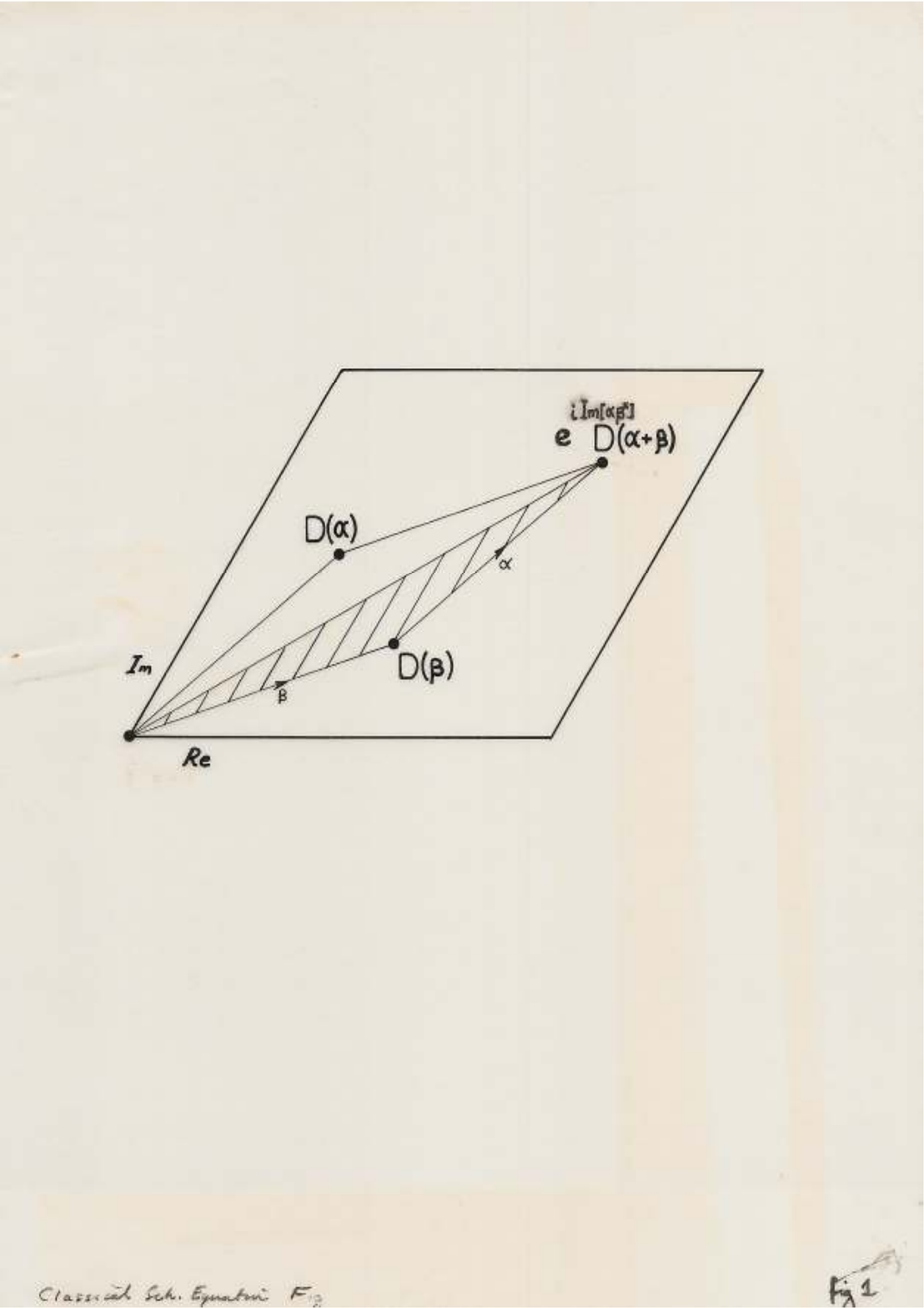}
\caption{The parameter space of Weyl operators is a plane, multiplication of
two of them, say $D(\alpha)$ and $D(\beta)$, leads to the new unitary
operator  $e^{i{\rm Im}[\alpha\beta^{*}]} D(\alpha + \beta)$, where the phase
factor derives from the ray character of this representation of the Abelian
group of translations upon the plane.\label{fig1}.}
\end{figure}

\begin{figure}[p]
\includegraphics[width=80mm]{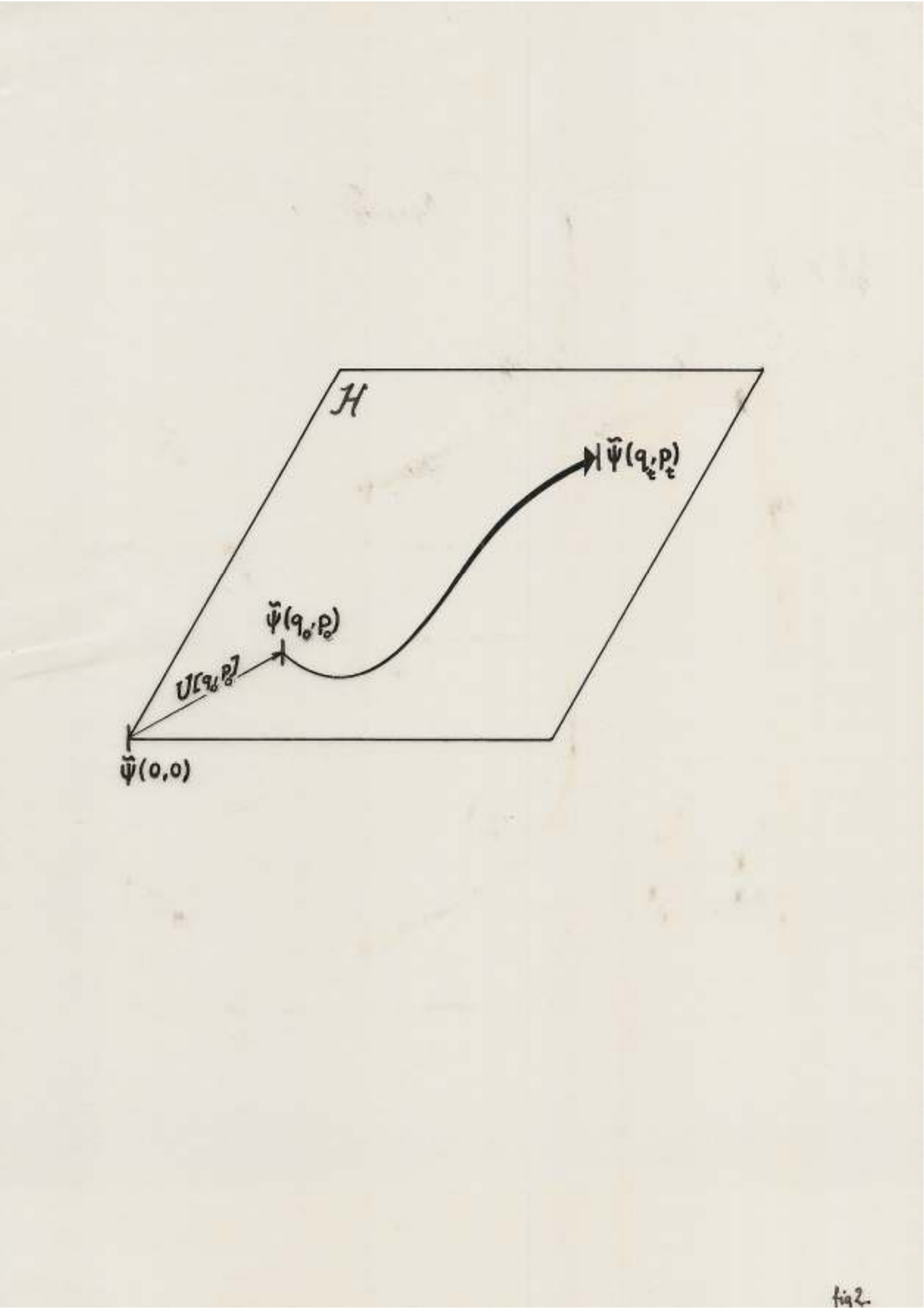}
\caption{Hilbert space $\cal H$ considered as the Weyl translates of a
special class of states $\tilde{{\psi}}(0,0)=\{\phi_{0}\}$, where $\phi_{0}$
runs through all states having both expectation values equal to zero. The
points $\tilde{{\psi}}(q,p)$ in this special quantal phase space thus
represent entire classes of wavefunctions which enjoy a one-one
correspondence with the phase space points of ordinary classical
mechanics.\label{fig2}}
\end{figure}

\begin{figure}[p]
\includegraphics[width=80mm]{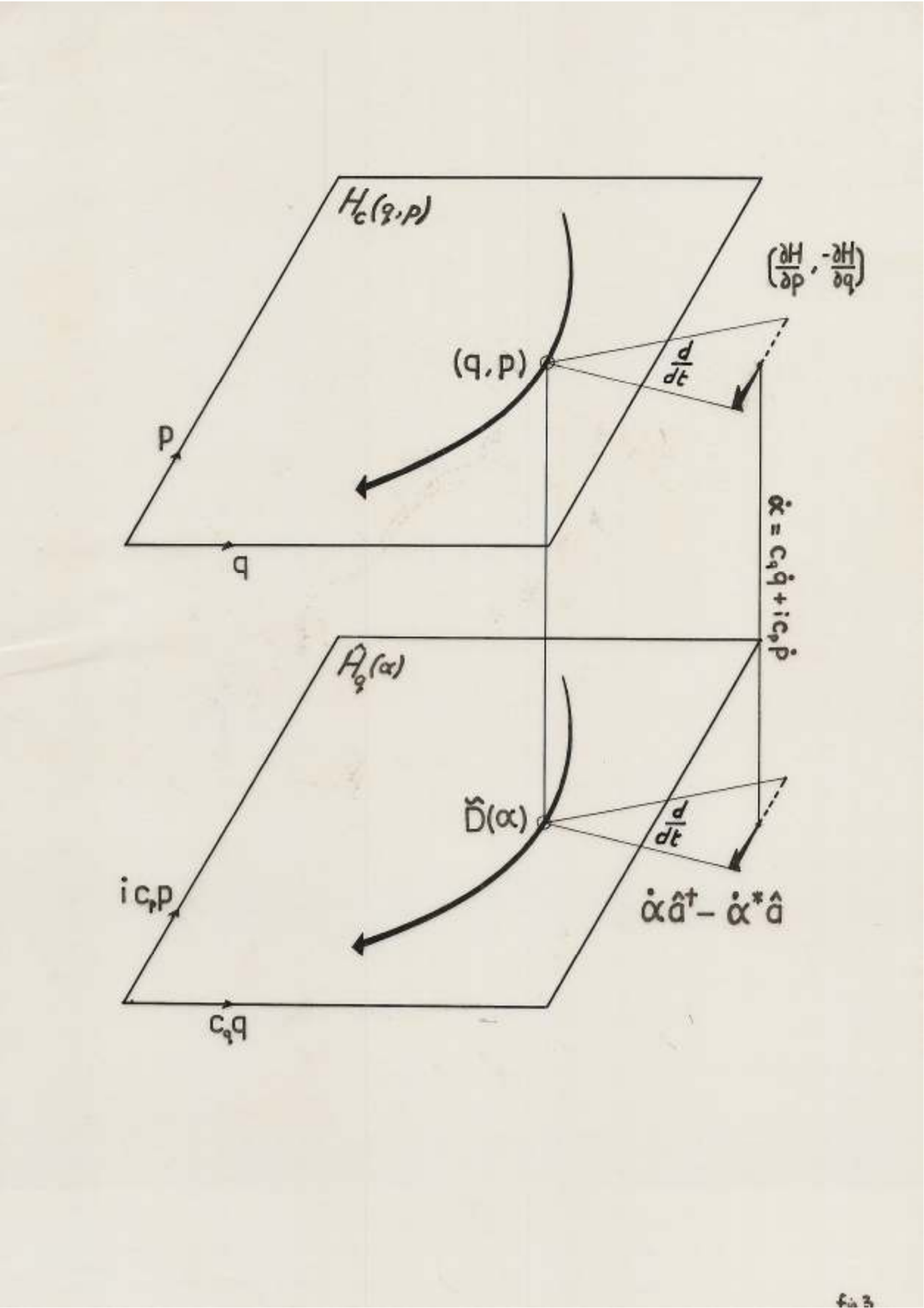}
\caption{The classical and quantal phase planes. In the upper classical plane
there is a trajectory $(q(t),p(t))$ being a solution to the Hamiltonian system
$H_{c}(q,p)$. In the lower quantal plane there is a Weyl operator trajectory 
$\tilde{D}(\alpha(t))$, which we fix up to an arbitrary time--dependent phase
via the projection $c_{q}q(t) + ic_{p}p(t) \mapsto \alpha(t)$. The choice
of $c_{q}$ and $c_{p}$ sets the scale relation between both planes, where
$\hbar = [2c_{q}c_{p}]^{-1}$.\label{fig3}}
\end{figure}

\begin{figure}[p]
\includegraphics[width=80mm]{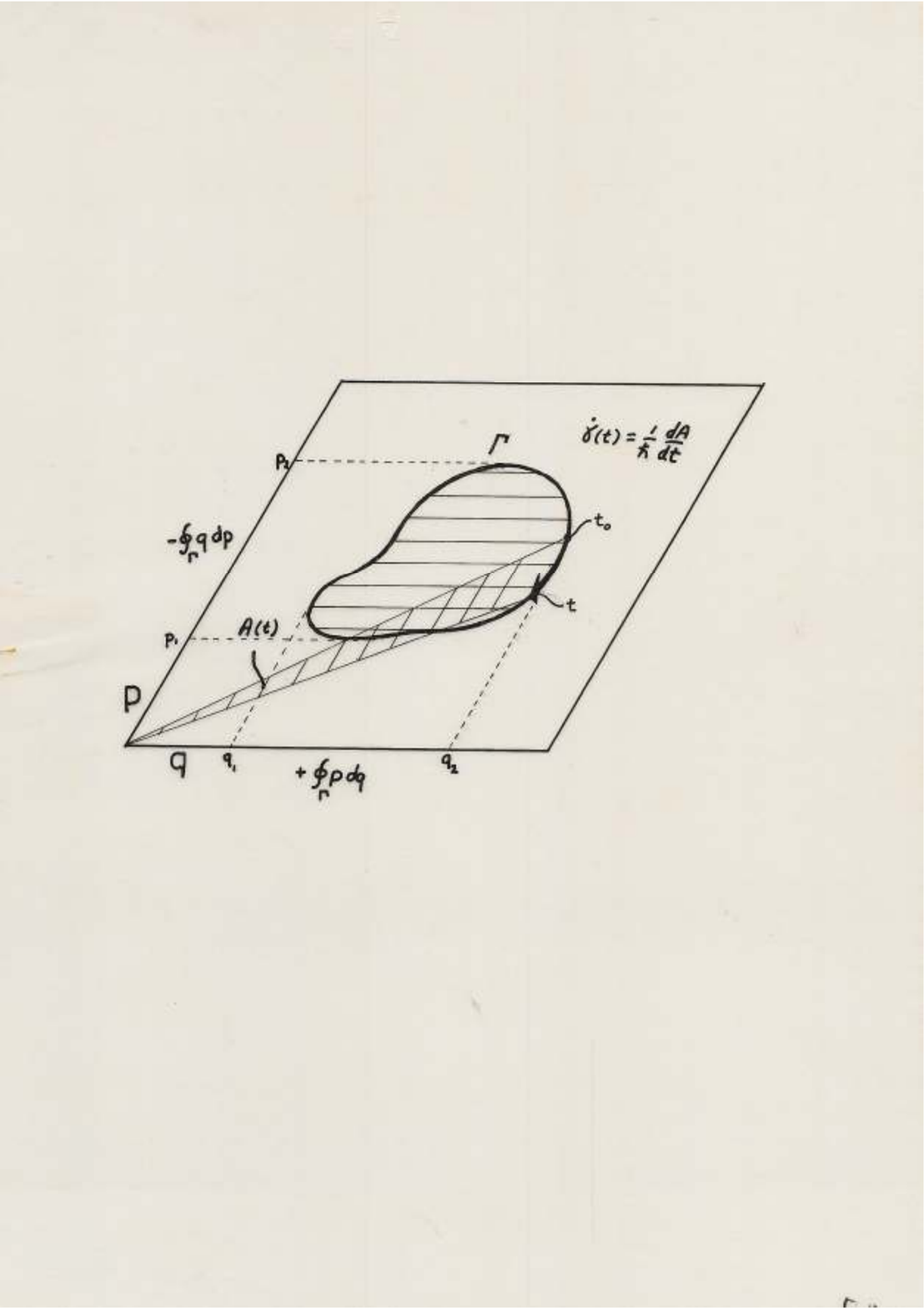}
\caption{The classical geometric phase embedded in ordinary classical phase
space. The form of $\gamma(t)$ is such that it equals the integrated
sectorial velocity, as measured from the origin. Upon closed circuits
$\Gamma$, the phase $\gamma(\Gamma)$ can be decomposed into the two
terms: $+1/2\hbar\oint_{\Gamma}p\,dq$ and $-1/2\hbar\oint_{\Gamma}q\,dp$.
Accounting for the sense of traversal, these terms are seen to be numerically
equal. Hence $\gamma(\Gamma) = +1/\hbar\oint_{\Gamma}p\,dq$.\label{fig4}}
\end{figure}


\begin{references}
\bibitem[31]{origin}There is no preferred origin in this space. The transformations $\hat{p}' \equiv U[q,p]^{\dagger} \hat{p} U[q,p]$ and $\hat{q}'\equiv U[q,p]^{\dagger} \hat{q} U[q,p]$  preserve the commutation relations, and thus the group algebra. Under this change the old origin at the unprimed class $\tilde{\psi}(0,0)$ shifts to the new origin at the primed class $\tilde{\psi}'(0,0)$, whose members are identical up to phase with those of the unprimed class $\tilde{\psi}(-q,-p)$.
\end{references}
\end{document}